\pdfoutput=1
\documentclass[preprint,showkeys,secnumarabic,amsfonts,showpacs,amsmath,amssymb]{revtex4}
\usepackage{epsfig}

\usepackage{euscript,graphicx,amssymb,subfigure}

\usepackage{amsfonts}
\usepackage{amsmath}
\usepackage{amssymb}
\usepackage{graphicx}
\usepackage{euscript}
\usepackage{color}
\usepackage{subfigure}

\newcommand{\goodgap}{\hspace{\subfigtopskip} \hspace{\subfigbottomskip}}

\def\Box{\hbox{$\rlap{$\sqcup$}\sqcap$}}

\begin{document}

\title{\bf Constraints on scalar-tensor theories from observations }
\author{Hossein Farajollahi}
\email{hosseinf@guilan.ac.ir}
\affiliation{$^1$Department of Physics, University of Guilan, Rasht, Iran}
\affiliation{$^2$ School of Physics, University of New South Wales, Sydney, NSW, 2052, Australia}

\author{Amin Salehi}
\affiliation{Department of Physics, University of Guilan, Rasht, Iran}
\author{Mohammad Nasiri}
\affiliation{Department of Physics, University of Guilan, Rasht, Iran}

\date{\today}

\begin{abstract}
We study the dynamical description of scalar-tensor gravity by performing the best-fit analysis for two cases of exponential and power-law form of the potential and scalar field function coupled to the curvature. The models are then tested against observational data. The results show that in both scenarios the Universe undergoes an acceleration expansion period and the geometrical equivalent of dark energy is associated with a time-dependent equation of state.

\end{abstract}



\maketitle

\section{Introduction}

Recent cosmological observations of high redshift type Ia
supernovae, the galaxy clusters survey \cite{Reiss}--\cite{Riess2}, Sloan digital sky survey ({\bf
SDSS})~\cite{Abazajian} and Chandra X--ray observatory~\cite{Allen} discover the universe accelerating expansion. Also Cosmic Microwave Background (CMB)
anisotropies observations \cite{Bennett} indicate universe flatness and that the total energy
density is close to one \cite{Spergel}. The observations determine basic cosmological parameters
with high precisions and also strongly indicate that the universe
presently is dominated by a smoothly distributed and slowly
varying dark energy (DE) component. A dynamical equation of state ( EoS) parameter that is related directly to the evolution of the energy density in an expanding universe can be taken as an appropriate parameter to interpret the universe acceleration \cite{Seljak}--\cite{Setare}.

Suggested by string theories \cite{Fujii} that, the scalar-tensor models provide the simplest model-independent description of unification theories which predict couplings between scalar fields and curvature.
 They have assumed a prominent role in cosmology since any unification scheme, such as supergravity in the weak
energy limit, or inflationary models such as chaotic inflation, seem to be supported by them \cite{capelo}. In addition, they derive the universe acceleration and possible phantom crossing \cite{Sahoo}--\cite{Carr}.

 In this paper, we study the dynamics of the scalar–-tensor theories by first best fitting the model with the observational data for distance modulus using $\chi^2$ statistical test. This allows us to observationally verify model before checking the results against any other experimental data. The layout of the paper is as follows. Section two recalls the basic dynamics of scalar-tensor theories. In Section three, in terms of new dynamical variables and for two power law and exponential forms of the functions $F(\phi)$ and $V(\phi)$ in the model, we numerically solve the equations. In section four, we best fit the model parameters and initial conditions against observational data for distance modulus. In section five we examine the model with two cosmological tests; the Eos parameter and the observational data for Hubble parameter. Finally we summarize and conclude in section six.

\section{THE MODEL}

A general action in four dimensions, where gravity is nonminimally coupled to a scalar field $\phi$, is given by
\begin{eqnarray}\label{ac1}
S=\frac{1}{16\pi G}\int[F(\phi)R-\frac{1}{2}\phi_{,\mu}\phi^{,\mu}+V(\phi)]\sqrt{-g}d^{4}x,
\end{eqnarray}
where R is the scalar curvature of $g_{\mu\nu}$, $g$ its determinant and $F(\phi)$ and $V(\phi)$ are two generic functions representing the coupling of the scalar field with geometry and its potential energy
density respectively. We chose $16\pi G=1$. The dynamics of the real scalar field $\phi$ depends a priori on the functions $F(\phi)$ and $V(\phi)$. By using the transformation $\phi=F(\varphi)$ and $\omega(\phi)=\frac{F(\varphi)}{2dF(\varphi)/d\varphi}$ in the action (\ref{ac1}), the Brans-Dicke theory can be recovered. Also, the standard Newton coupling is recovered in the limit $F(\phi)\rightarrow -\frac{1}{2}$.

The action \ref{ac1} with negative sign of the kinetic term seems to contain ghost fields for certain
values of the non-minimal coupling functions $F(\phi)$. The presence of such
fields would explain how the dominant energy condition is violated in the
model, as is required to give rise to phantom behavior. Ghost fields in
the theory lead to instability of the vacuum and the model may suffers from catastrophic theoretical instabilities. However, as follows by assuming power law or exponential forms of $F(\phi)$ and best fitting the model, the function is always positive as also required for the conformal transformation to be well defined.

The field equations can be derived by varying the action (\ref{ac1}) with respect to $g_{\mu\nu}$
\begin{eqnarray}\label{ac2}
F(\phi)(R_{\mu\nu}-\frac{1}{2}g_{\mu\nu}R)=T_{\mu\nu}^{(\phi)},
\end{eqnarray}
where
\begin{eqnarray}
T_{\mu\nu}^{(\phi)}=\frac{1}{2}\phi_{;\mu}\phi_{;\nu}-\frac{1}{4}g_{\mu\nu}\phi_{;\alpha}\phi_{;\alpha}
-g_{\mu\nu}\Box F(\phi)+F(\phi)_{;\mu\nu}+g_{\mu\nu}V(\phi).
\end{eqnarray}
In addition, variation with respect to the scalar field $\phi$  gives the klein-Gordan equation,
\begin{eqnarray}
\Box \phi+R(\frac{dF}{d\phi})+V(\phi)=0.
\end{eqnarray}
Assuming a spatially homogeneous scalar field $\phi(t)$, and in background of a flat FRW metric, the field equations become,
\begin{eqnarray}
3H^{^{2}}F&=&-3H\dot{F}+\frac{\dot{\phi}^{2}}{4}+\frac{V}{2},\label{fried1}\\
-2\dot{H}F&=&3H^{2}F+2H\dot{F}+\ddot{F}+\frac{\dot{\phi}^{2}}{4}
+\frac{V}{2},\label{fried2}\\
\ddot{\phi}\dot{\phi}&+&3H\dot{\phi}^2=6(\dot{H}+2H^{2})\dot{F}-\dot{V},\label{phiequation}
\end{eqnarray}
where equation (\ref{fried1}) is the energy constraint corresponding to the (0,0)-Einstein equation. In comparison with the standard cosmological model, one can find an effective EoS for the model as $p_{eff}\equiv\omega_{eff}\rho_{eff}$, where $\rho_{eff}$ and $p_{eff}$ are respectively equivalent to the right hand side of the equation (\ref{fried1}) and (\ref{fried2}). In general, the system of nonlinear second order differential equations (\ref{fried1})-(\ref{phiequation}) has no known analytic solution. However, in the following by introducing new dimensionless dynamical variables, we replace the above equations with a system of nonlinear first order differential equations which is capable to solve numerically.

\section{Dynamical system}

In this section, we study the structure of the dynamical system by introducing the following dimensionless variables,
\begin{eqnarray}\label{defin}
\Omega_{F}={-\frac{\dot{F}}{F H}},         \Omega_{\dot{\phi}}={\frac{\dot{\phi}^{2}}{12FH^{2}}},              \Omega_{V}={\frac{V}{6FH^{2}}}.
\end{eqnarray}
In two different scenarios for the scalar field function $F(\phi)$ and potential $V(\phi)$ we investigate the dynamics of the model. The cosmological models with such functions have been known lead to interesting physics
in a variety of context, ranging from existence of accelerated expansions \cite{Halliwell} to cosmological
scaling solutions \cite{Ratra}--\cite{Yokoyama}.

{\bf Power law $F(\phi)$ and $V(\phi)$ }

Using equations (\ref{fried1})-(\ref{phiequation}), and power low forms for $F(\phi)\equiv\phi^{\alpha}$ and $V(\phi)\equiv \phi^{\beta}$, the dynamical equations of the new variables become,
\begin{eqnarray}
\frac{d\Omega_{F}}{dN}&=&-\frac{\alpha\ddot{\phi}}{\phi H^{2}}+\frac{\Omega_{F}^{2}}{\alpha}-\Omega_{F}\frac{\dot{H}}{H^{2}},\label{omegaf}\\
\frac{d\Omega_{\dot{\phi}}}{dN}&=&-\frac{\alpha\ddot{\phi}}{\phi H^{2}}\frac{2\Omega_{\dot{\phi}}}{\Omega_{F}}+\Omega_{F} \Omega_{\dot{\phi}}-2\Omega_{\dot{\phi}}\frac{\dot{H}}{H^{2}},\label{omegaphi}\\
\frac{d\Omega_{V}}{dN}&=&\frac{\alpha\ddot{\phi}}{\phi H^{2}}(1+\frac{2\Omega_{\dot{\phi}}}{\Omega_{F}})-\Omega_{F}\Omega_{\dot{\phi}}
-\frac{\Omega_{F}^{2}}{\alpha}+(\Omega_{\dot{\phi}}+\Omega_{F})\frac{\dot{H}}{H^{2}},\label{omegav}
\end{eqnarray}
where $N = ln (a)$ and also
\begin{eqnarray}
\frac{\dot{H}}{H^{2}}&=&\frac{-2\Omega_{\dot{\phi}}}{4\Omega_{\dot{\phi}}+\Omega_{F}^{2}}(3+\Omega_{F}
+\frac{\alpha-1}{\alpha}\Omega_{F}^{2}+3\Omega_{\dot{\phi}}+3 \Omega_{V}+\frac{\Omega_{F}^{2}}{\Omega_{\dot{\phi}}}-\frac{\beta \Omega_{V}\Omega_{F}^{2}}{2\alpha \Omega_{\dot{\phi}}}),\\
\frac{\alpha\ddot{\phi}}{\phi H^{2}}&=&3\Omega_{F} +\frac{\Omega_{F}^{2}}{\Omega_{\dot{\phi}}}+\frac{\Omega_{F}^{2}}{2\Omega_{\dot{\phi}}}\frac{\dot{H}}{H^{2}}
-\frac{\beta \Omega_{V}\Omega_{F}^{2}}{2\alpha \Omega_{\dot{\phi}}}.
\end{eqnarray}
By using the Friedmann constraint equation (\ref{fried1}) which now becomes,
\begin{eqnarray}\label{constraint2}
\Omega_{V}+\Omega_{F}+\Omega_{\dot{\phi}}=1.
\end{eqnarray}
the equations (\ref{omegaf})-(\ref{omegav}) reduce to two differential equations for $\Omega_{F} $ and $\Omega_{\dot{\phi}}$. However, the differential equations are long and tedius and so are not presented in here.

{\bf Exponential $F(\phi)$ and $V(\phi)$}

Similarly, using equations (\ref{fried1})-(\ref{phiequation}), and exponential forms of $F(\phi)\equiv e^{\alpha\phi}$ and $V(\phi)\equiv e^{\beta\phi}$ the evolution equations become,
\begin{eqnarray}
\frac{d\Omega_{F}}{dN}&=&-\frac{\alpha\ddot{\phi}}{ H^{2}}-\Omega_{F}\frac{\dot{H}}{H^{2}},\label{omegaf1}\\
\frac{d\Omega_{\dot{\phi}}}{dN}&=&-\frac{\alpha\ddot{\phi}}{ H^{2}}\frac{2\Omega_{\dot{\phi}}}{\Omega_{F}}+\Omega_{F} \Omega_{\dot{\phi}}-2\Omega_{\dot{\phi}}\frac{\dot{H}}{H^{2}}\label{omegaphi1}\\
\frac{d\Omega_{V}}{dN}&=&\frac{\alpha\ddot{\phi}}{ H^{2}}(1+\frac{2\Omega_{\dot{\phi}}}{\Omega_{F}})-\Omega_{F}\Omega_{\dot{\phi}}+(2\Omega_{\dot{\phi}}
+\Omega_{F})\frac{\dot{H}}{H^{2}},\label{omegav1}
\end{eqnarray}
where
\begin{eqnarray}
\frac{\dot{H}}{H^{2}}&=&\frac{-2\Omega_{\dot{\phi}}}{4\Omega_{\dot{\phi}}+\Omega_{F}^{2}}(3+\Omega_{F}
+\Omega_{F}^{2}+3\Omega_{\dot{\phi}}+3 \Omega_{V}+\frac{\Omega_{F}^{2}}{\Omega_{\dot{\phi}}}-\frac{\beta \Omega_{V}\Omega_{F}^{2}}{2\alpha \Omega_{\dot{\phi}}})\\
\frac{\alpha\ddot{\phi}}{ H^{2}}&=&3\Omega_{F} +\frac{\Omega_{F}^{2}}{\Omega_{\dot{\phi}}}+\frac{\Omega_{F}^{2}}{2\Omega_{\dot{\phi}}}\frac{\dot{H}}{H^{2}}
-\frac{\beta \Omega_{V}\Omega_{F}^{2}}{2\alpha \Omega_{\dot{\phi}}}.
\end{eqnarray}
Together with the Friedmann constraint equation (\ref{constraint2}) for this scenario similar to the previous case the equations (\ref{omegaf1})-(\ref{omegav1}) reduce to two differential equations for $\Omega_{F} $ and $\Omega_{\dot{\phi}}$. Again, since the differential equations are long and messy, we omit them.

Still, solving analytically the set of non-linear first-order differential equations in both cases of exponential or power-law $F(\phi)$ and $V(\phi)$ is difficult. So in the following we present a numerical solution for them. In addition, we simultaneously best fit the model parameters  $\alpha$, $\beta$ and initial conditions $\Omega_{F}(0), \Omega_{\dot{\phi}}(0)$, $H(0)$ with the observational data using the $\chi^2$ method when solving the equations for two exponential and power-law scenarios. The advantage of simultaneously solving the system of equations and best fitting is that the solutions are  physically meaningful and observationally favored.

\section{Best-fitting and cosmological constraints}

In the following we numerically solve the field equations and also best-fit our model parameters with three sets of observational data; i) The Sne Ia dataset, ii) Cosmic Microwave Background (CMB) data, iii) Baryon Acoustic Oscillations (BAO) data.

{\bf The Sne Ia dataset}

The difference between the absolute and
apparent luminosity of a distance object is given by, $\mu(z) = 25 + 5\log_{10}d_L(z)$ where the Luminosity distance quantity, $d_L(z)$ is given by
\begin{equation}\label{dl}
d_{L}(z)=(1+z)\int_0^z{-\frac{dz'}{H(z')}}.
 \end{equation}
By numerical techniques, we solve the system of dynamical equations for $\Omega_{F} $ and $\Omega_{\dot{\phi}}$ in both power law and exponential cases. In addition, to best fit the model parameters and initial conditions, two auxiliary equations for the luminosity distance and the hubble parameter are employed:
\begin{eqnarray}
 \frac{dH}{dN}=H(-\frac{\dot{H}}{H^{2}}),
 \end{eqnarray}
\begin{eqnarray}
\frac{d(d_{L})}{dN}=-(d_{L}+\frac{e^{-2N}}{H}).
\end{eqnarray}
To best fit the model for the parameters $\alpha$, $\beta$ and the initial conditions $\Omega_{F}(0)$, $\Omega_{\dot{\phi}}(0)$, $H(0)$ with the most recent observational data, the Type Ia supernovea (SNe Ia), we employe the $\chi^2$ method. We constrain the parameters including the initial conditions by minimizing the $\chi^2$ function given as
\begin{equation}\label{chi2}
 \chi^2_{SNe} ( \alpha,\beta, \Omega_{F}(0), \Omega_{\dot{\phi}}(0), H(0))=\sum_{i=1}^{557}\frac{[\mu_i^{the}(z_i|\alpha,\beta, \Omega_{F}(0),\Omega_{\dot{\phi}}(0), H(0)) - \mu_i^{obs}]^2}{\sigma_i^2},
\end{equation}
where the sum is over the SNe Ia data. In relation (\ref{chi2}), $\mu_i^{the}$ and $\mu_i^{obs}$ are the distance modulus parameters obtained from our model and observation, respectively, and $\sigma$ is the estimated error of the $\mu_i^{obs}$.

{\bf CMB data}

For CMB data, the CMB shift parameter R, given by\cite{bond},\cite{wang}\\
\begin{eqnarray}
R=\Omega_{m0}^{\frac{1}2{}}\int_{0}^{z_{rec}}\frac{dz}{E(z)}
\end{eqnarray}
where $E(Z)=\frac{H(z)}{H_{0}}$ is the redshift of recombination $z_{rec} = 1090 $\cite{Komatsu}. The parameter R ties up the angular diameter
distance to the last scattering surface, the comoving size of the sound horizon at $z = 1091.3$ and the angular scale
of the first acoustic peak in CMB power spectrum of temperature fluctuations\cite{bond},\cite{wang}. The updated value of R from WMAP5 is $R_{obs} = 1.710 \pm 0.019 $ \cite{Komatsu}. It should be noted that, apart
from the SNIa and BAO data, the R parameter can provide information about the universe at very high
redshift. The $\chi_{CMB}^{2}$ for the CMB data is\\
\begin{eqnarray}
\chi_{CMB}^{2}=\frac{(R-R_{obs})^{2}}{\sigma_{R}}
\end{eqnarray}
where the corresponding $1\sigma$ errors is $\sigma_{R}$ = 0.019.\\

{\bf BAO data}

For BAO data, from the measurement of the BAO peak in the distribution
of SDSS luminous red galaxies, we define parameter A as \cite{Eisenstein} \\
\begin{eqnarray}
A=\Omega_{m0}^{\frac{1}{2}}E(z_{b})^\frac{-1}{3}[\frac{1}{z_{b}}\int_{0}^{z_{b}}\frac{dz}{E(z)}]^{\frac{2}{3}}
\end{eqnarray}
where $z_{b} = 0.35$. The SDSS BAO measurement \cite{Eisenstein} gives $A_{obs} = 0.469 (n_s/0.98).0.35 ± 0.017$, where the
scalar spectral index is taken to be $n_{s} = 0.960 $as measured by WMAP5 \cite{Komatsu}. The parameter A
is nearly model-independent and imposes robust constraint as complement to SNIa data. The $\chi^{2}$ for the
BAO data is\\
\begin{eqnarray}
\chi_{BAO}^{2}=\frac{(A-A_{obs})^{2}}{\sigma_{A}}
\end{eqnarray}
where the corresponding $1\sigma$ errors is $\sigma_{A} = 0.017$. Table I shows the best fitted model parameters and initial conditions in both power law and exponential cases.
\begin{table}[ht]
\caption{Best-fitted model parameters} % % title of Table
\centering % used for centering table
\begin{tabular}{c c c c c c c c} % centered columns (7 columns)
\hline %inserts double horizontal lines
Cosmological test &model  &  $\alpha$  &  $\beta$ \ & $\Omega_F(0)$\ & $\Omega_{\dot{\phi}}(0)$\ & $H(0)$ \ & $\chi^2_{min}$\\ [2ex] % inserts table
%heading
\hline % inserts single horizontal line
$SNIea $&power law&$-1.25$  & $4.44$ \ & $1.6$\ & $0.6$\ & $0.694$ \ & $544.6282552$ \\
$SNIea+CMB+BAO $ & power law & $-1.5$  & $5.95$ \ & $1.6$\ & $0.6$\ & $0.720$ \ & $683.5751181$ \\
$SNIea$ &  exponential&$-1.92$  & $5.88$ \ & $1.6$\ & $0.6$\ & $0.698$ \ & $542.9071824$ \\
$SNIea+CMB+BAO $& exponential & $-1.68$  & $6.7$ \ & $1.6$\ & $0.6$\ & $0.722$ \ & $675.4777593$ \\
\hline %inserts single line
\end{tabular}
\label{table:1} % is used to refer this table in the text
\end{table}\

Figs. 1-4 shows the constraints on the model parameters $\alpha$, $\beta$ and the initial conditions for $\Omega_F(0)$, $\Omega_{\dot{\phi}}(0)$ and $H(0)$ at the $68.3\%$, $95.4\%$ and $99.7\%$ confidence levels in both cases of power law and exponential functions.

\begin{figure*}
\centering
\subfigure{\includegraphics[width=7cm]{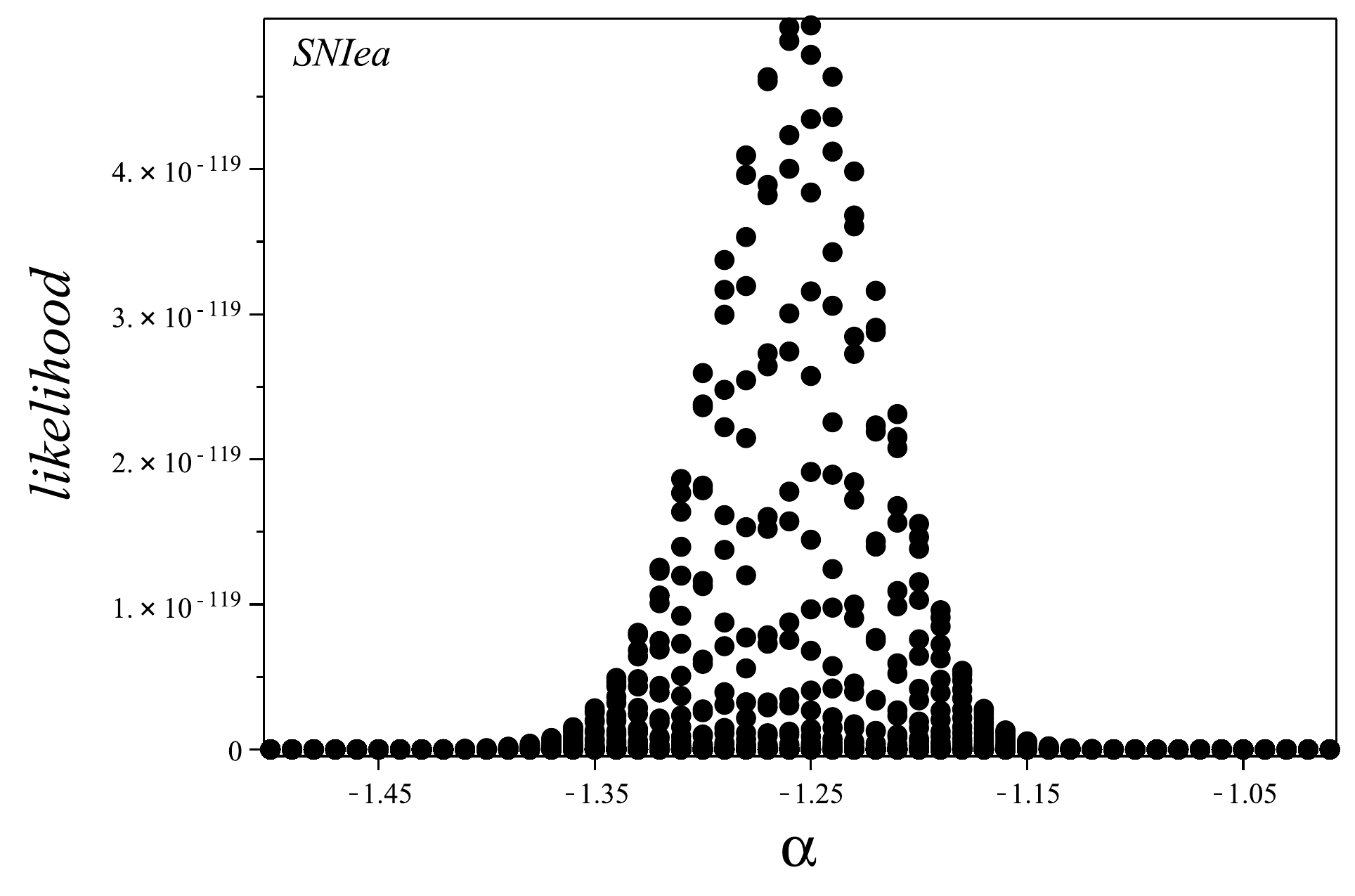}} \goodgap
\subfigure{\includegraphics[width=7cm]{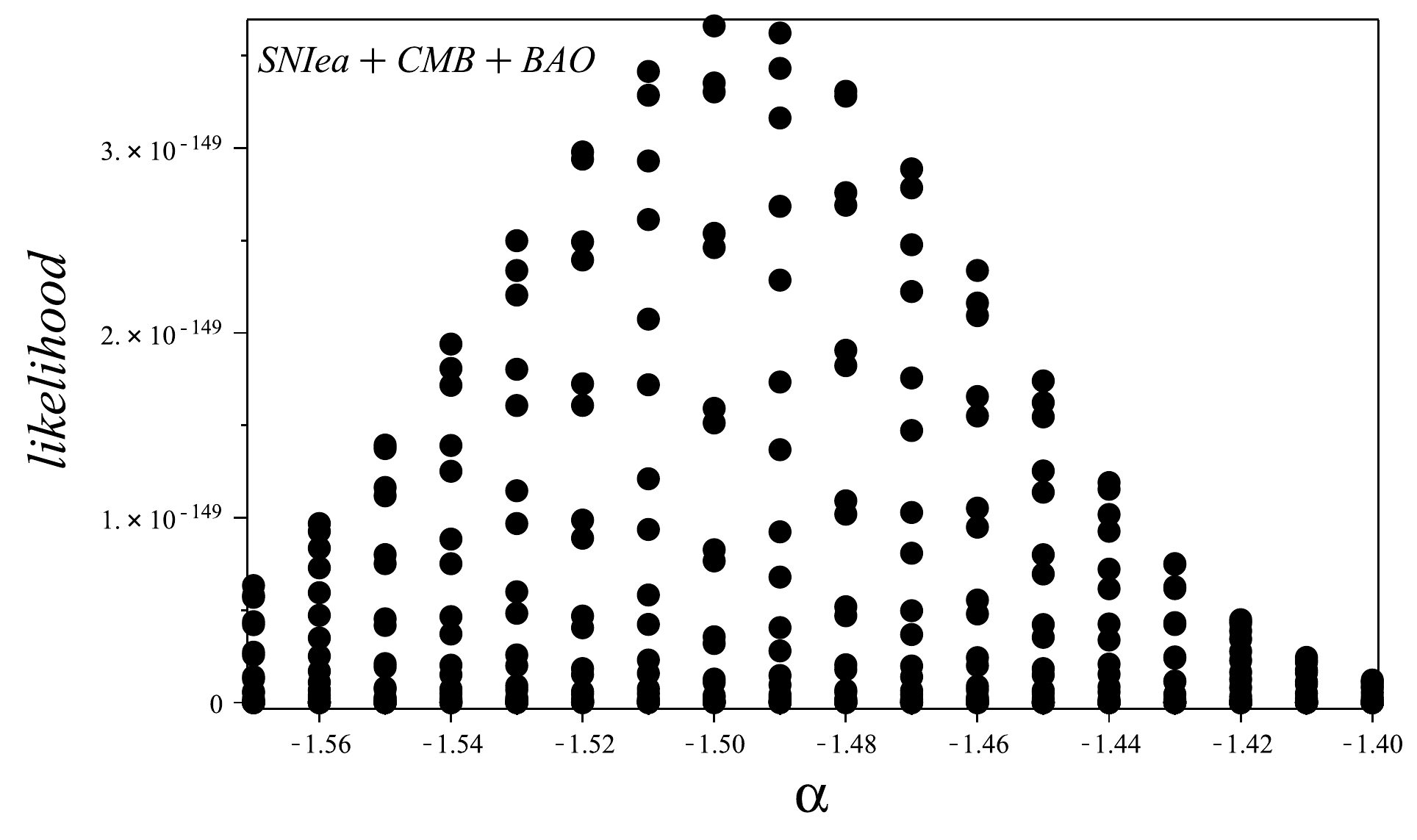}} \goodgap\\
\subfigure{\includegraphics[width=7cm]{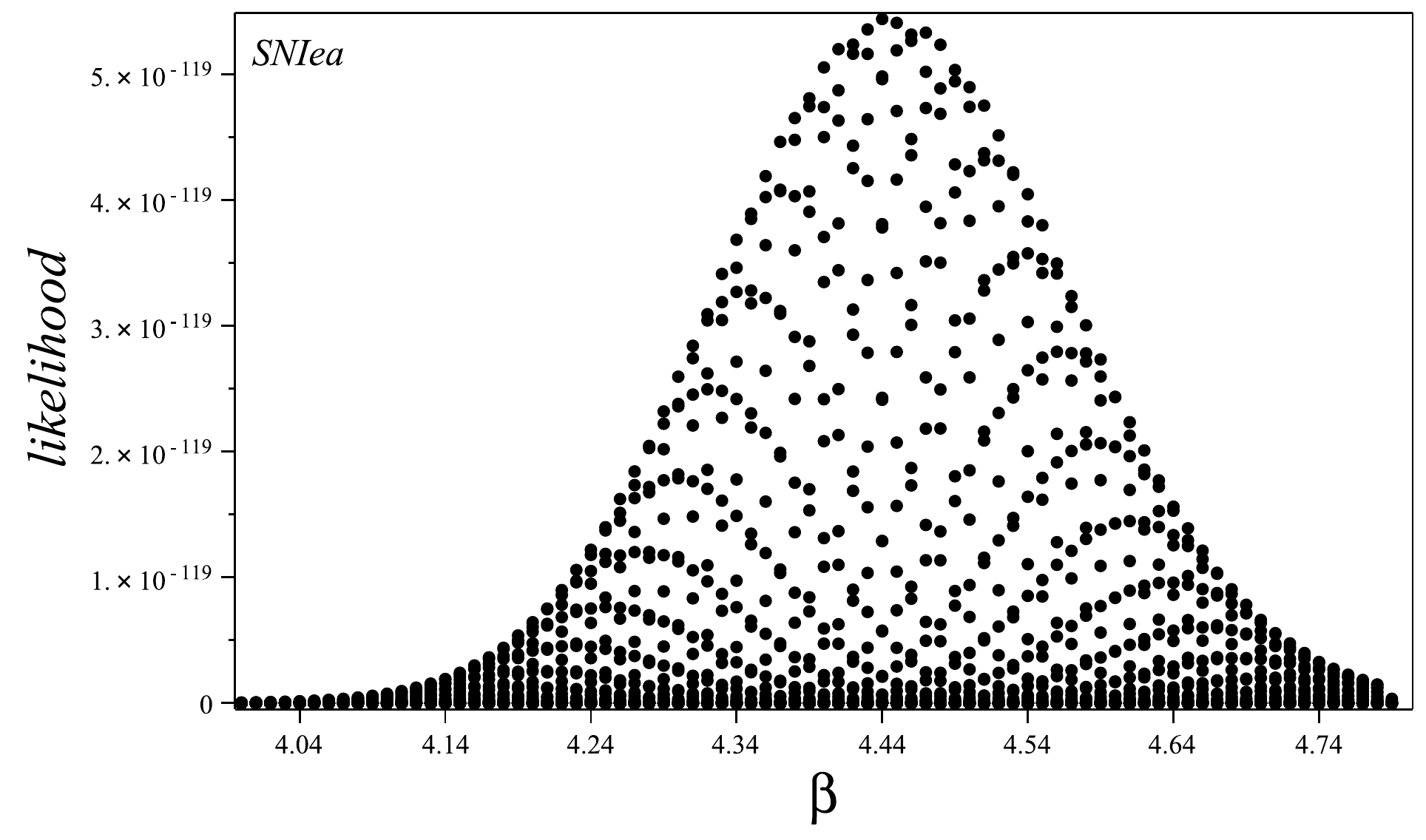}} \goodgap
\subfigure{\includegraphics[width=7cm]{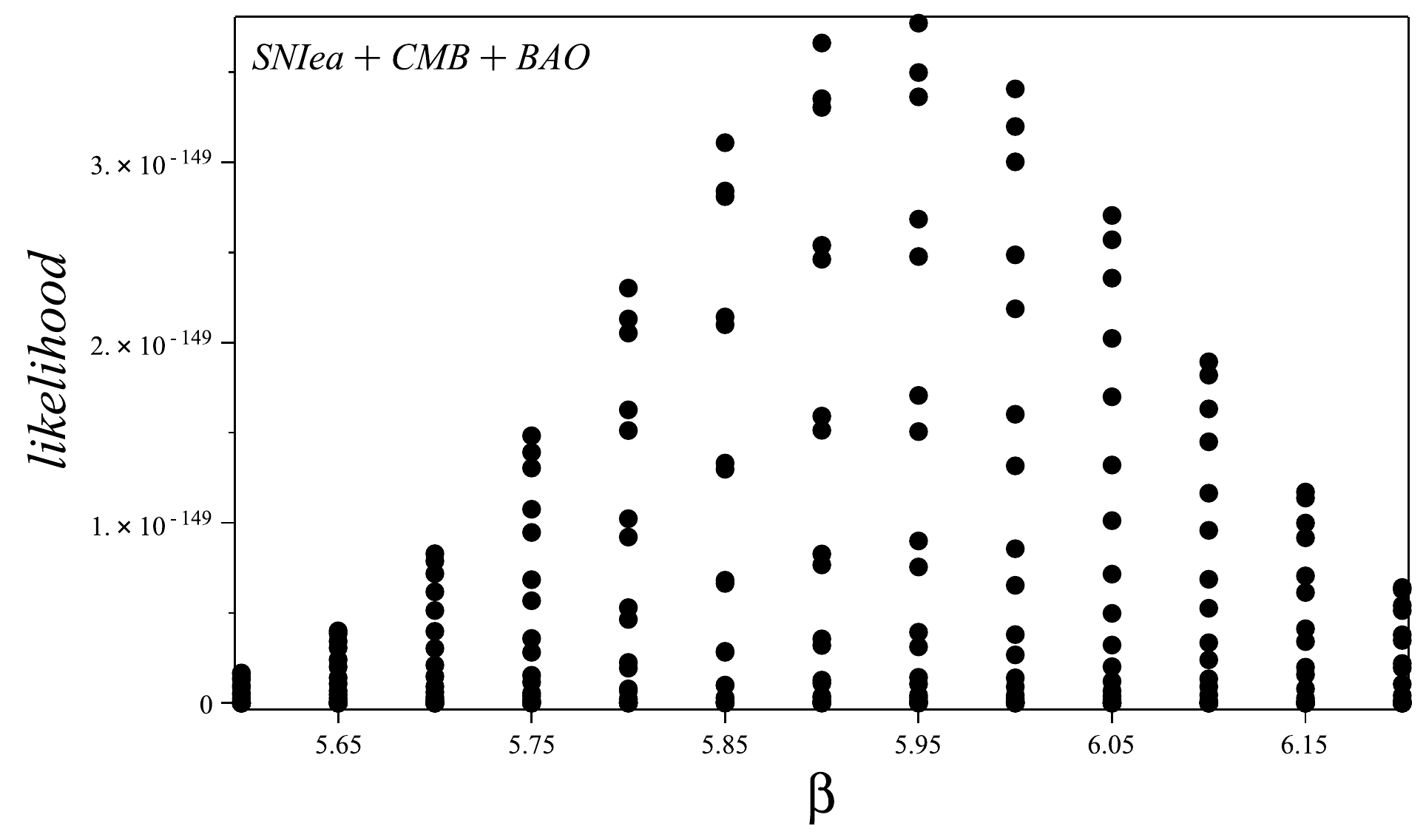}} \goodgap\\
\subfigure{\includegraphics[width=7cm]{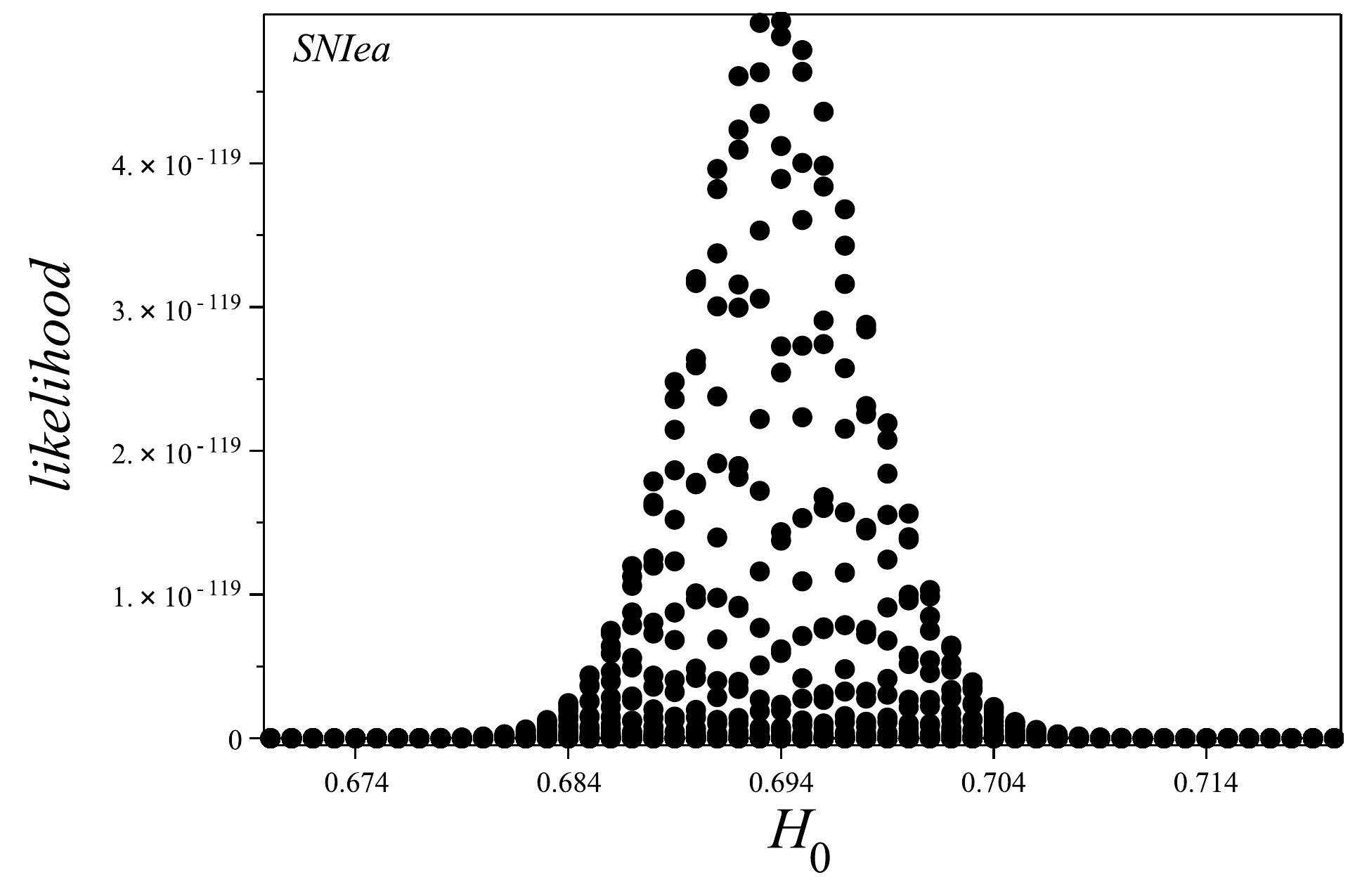}} \goodgap
\subfigure{\includegraphics[width=7cm]{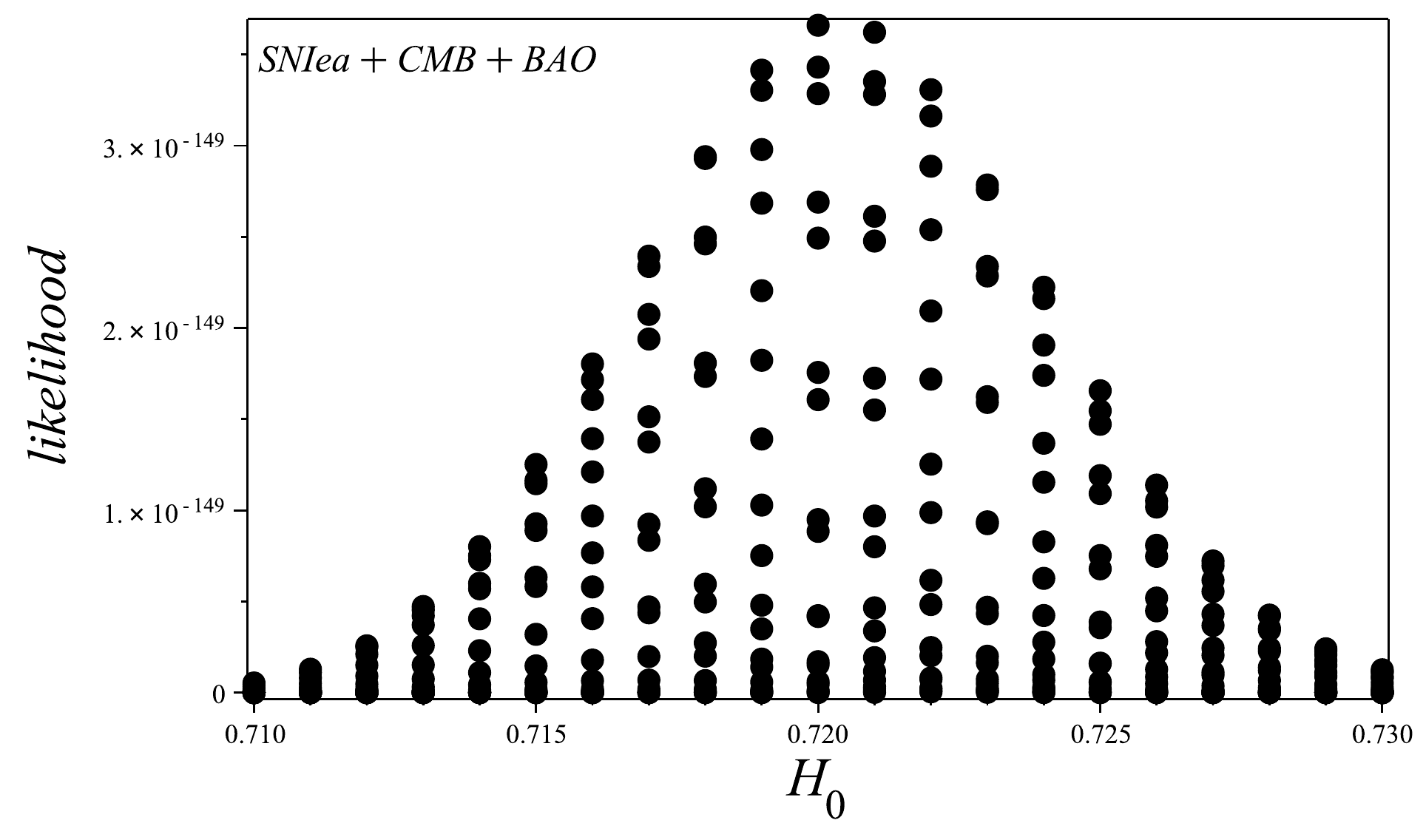}} \goodgap\\
\caption{The best-fitted one dimension likelihood and confidence level for $\alpha$, $\beta$ and $H_{0}$ for exponential functions}
\label{fig: clplots}
\end{figure*}
\begin{figure*}
\centering
\subfigure{\includegraphics[width=7cm]{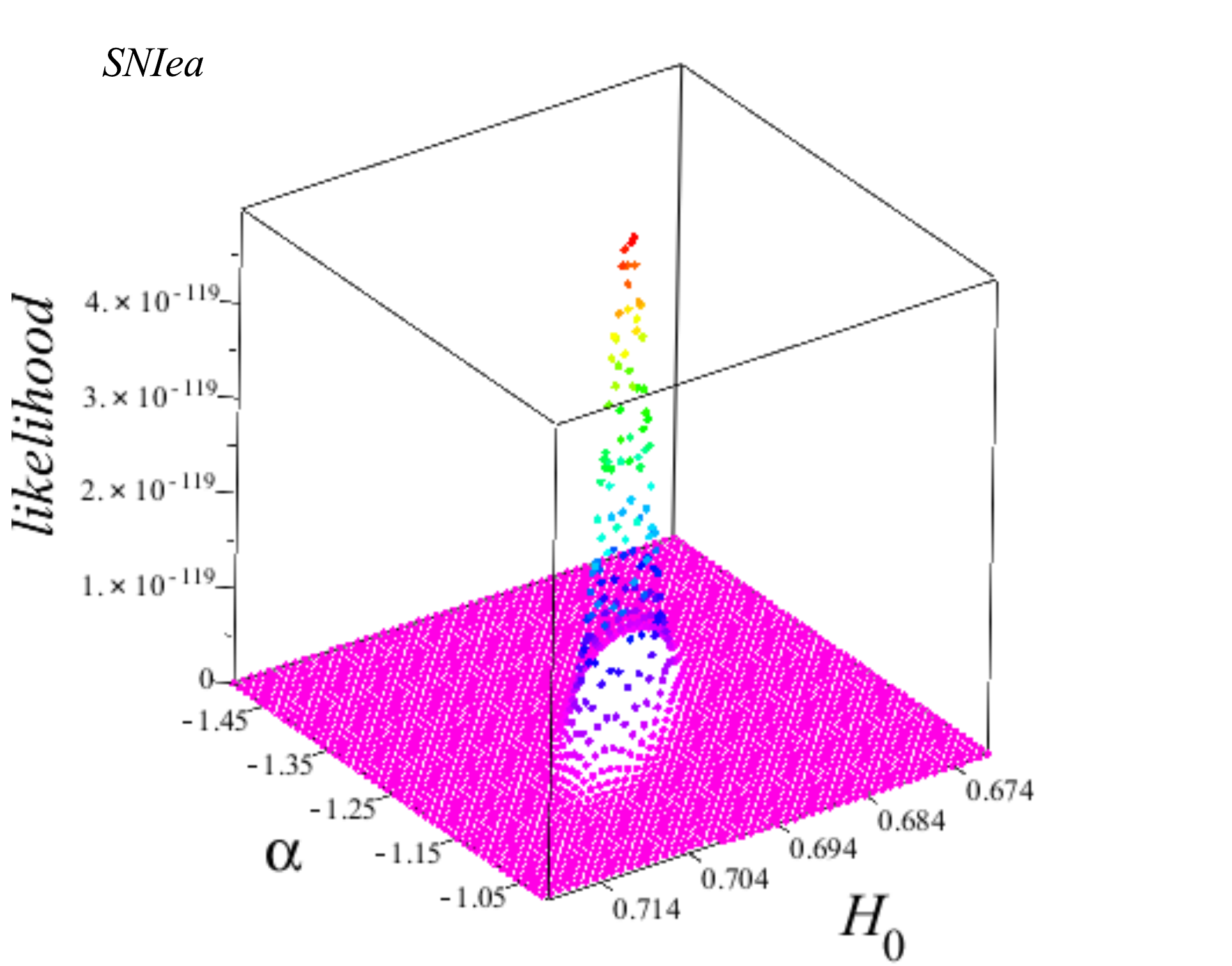}} \goodgap
\subfigure{\includegraphics[width=7cm]{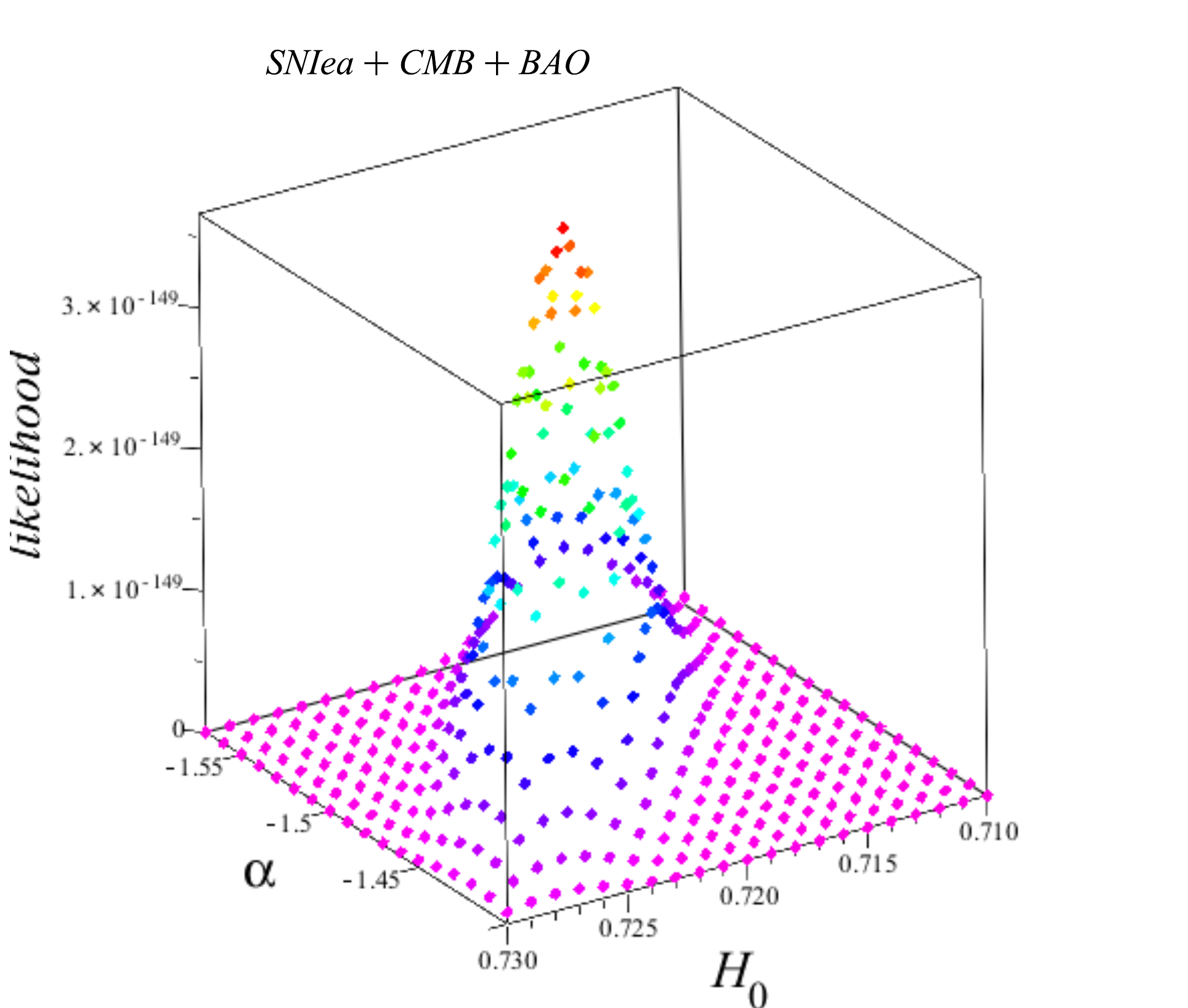}} \goodgap\\
\subfigure{\includegraphics[width=7cm]{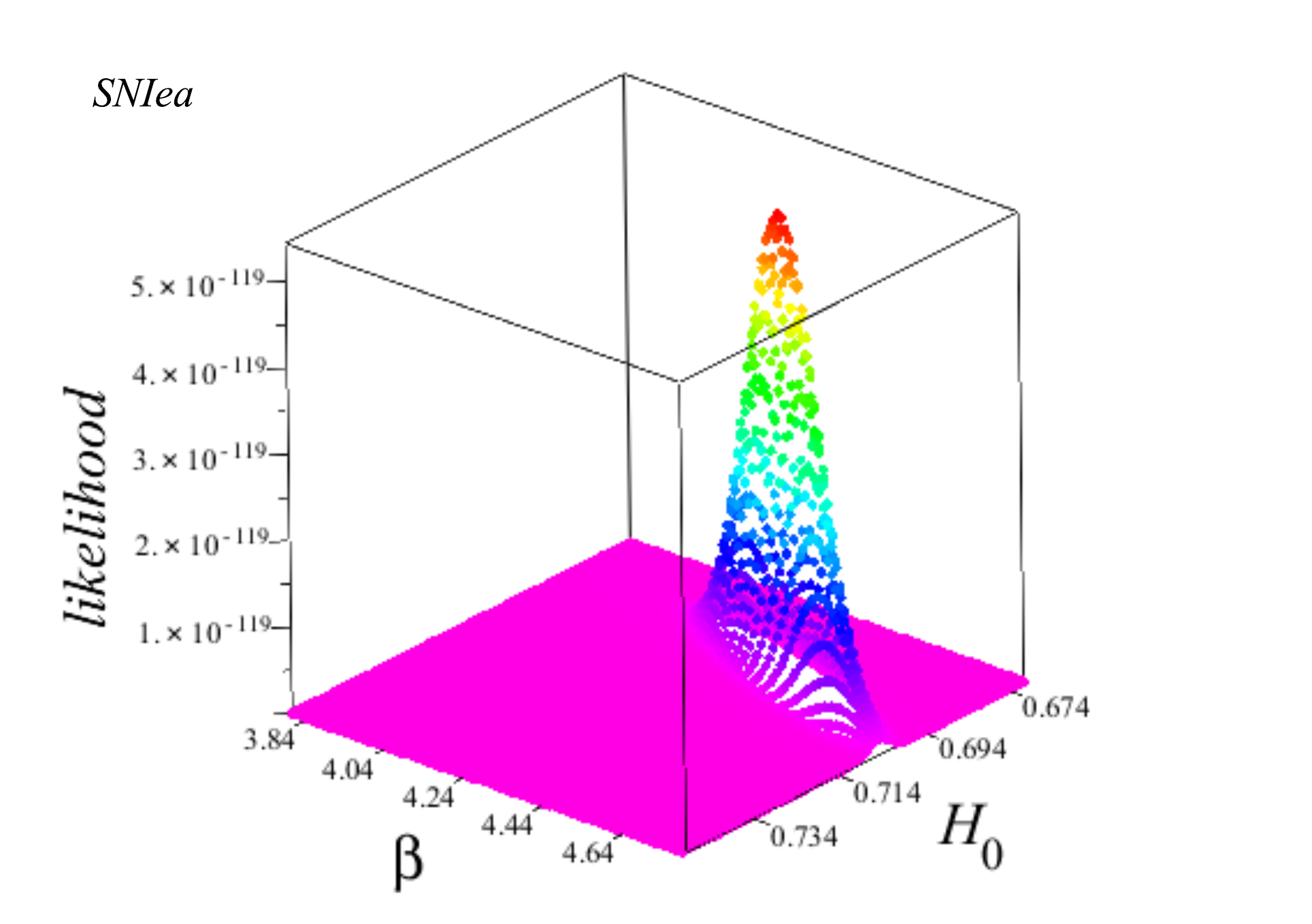}} \goodgap
\subfigure{\includegraphics[width=7cm]{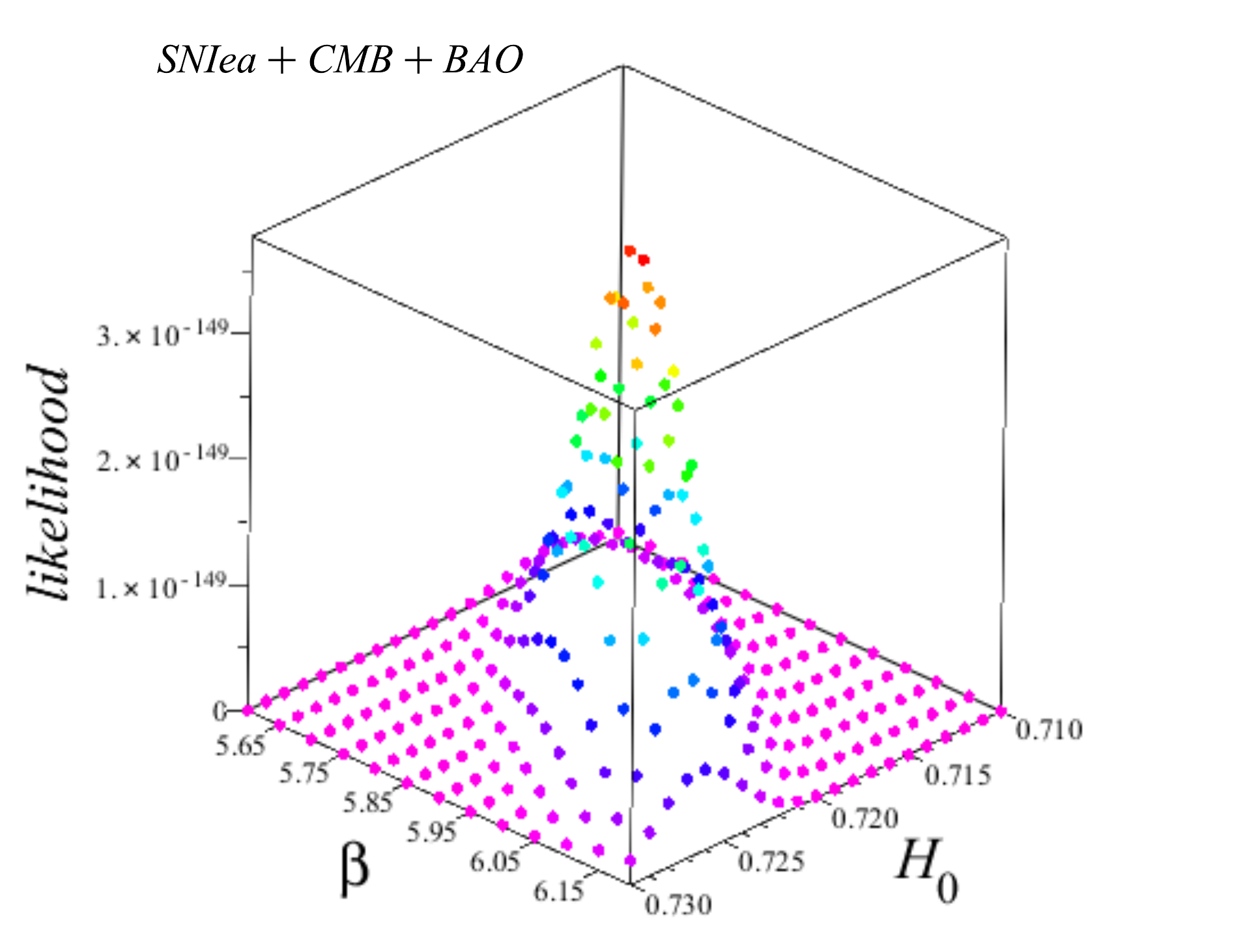}} \goodgap\\
\subfigure{\includegraphics[width=7cm]{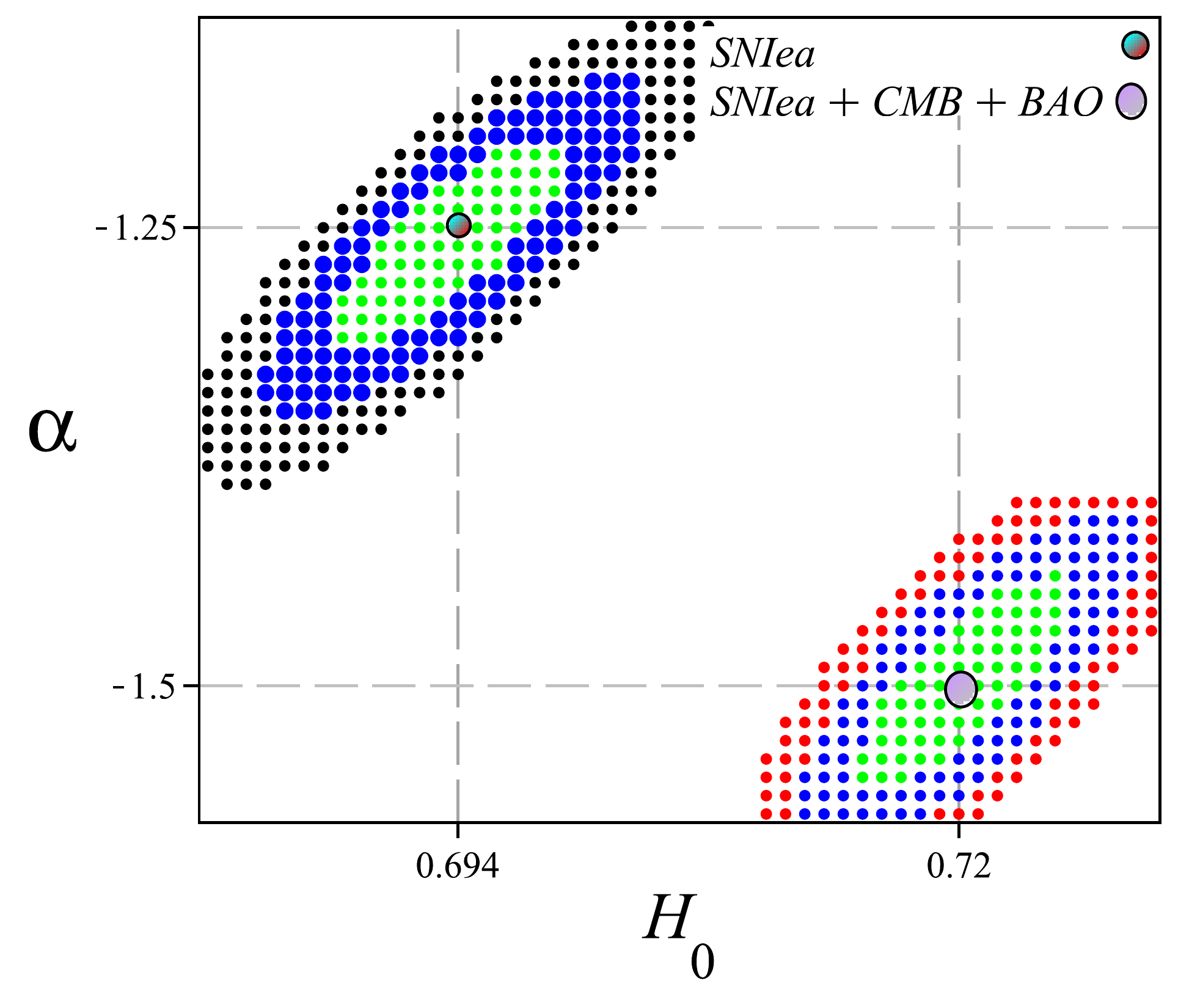}} \goodgap
\subfigure{\includegraphics[width=7cm]{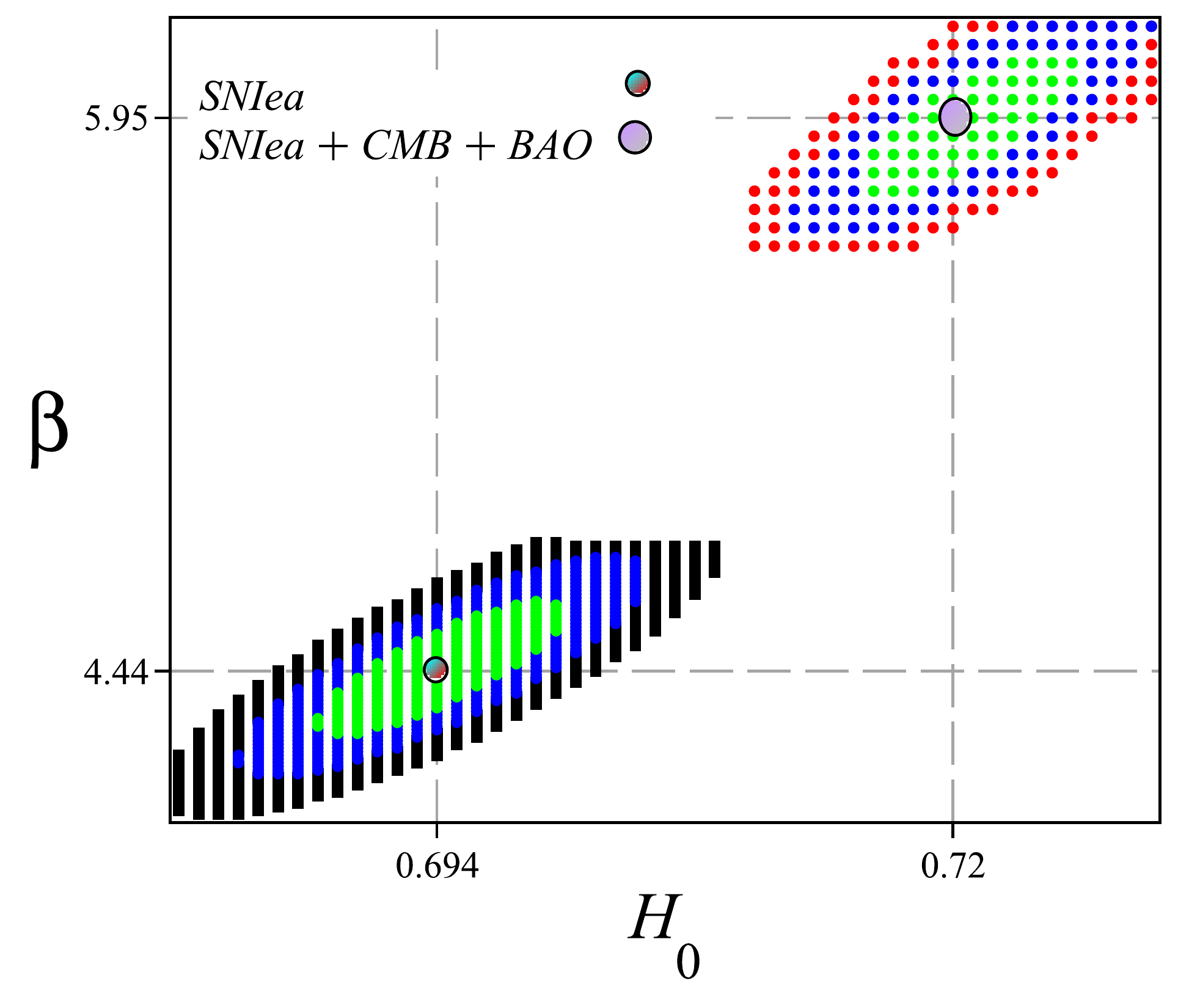}} \goodgap\\
\caption{The best-fitted two dimension likelihood and confidence level for $\alpha$, $\beta$ and $H_{0}$ for exponential functions}
\label{fig: clplots}
\end{figure*}

\begin{figure*}
\centering
\subfigure{\includegraphics[width=7cm]{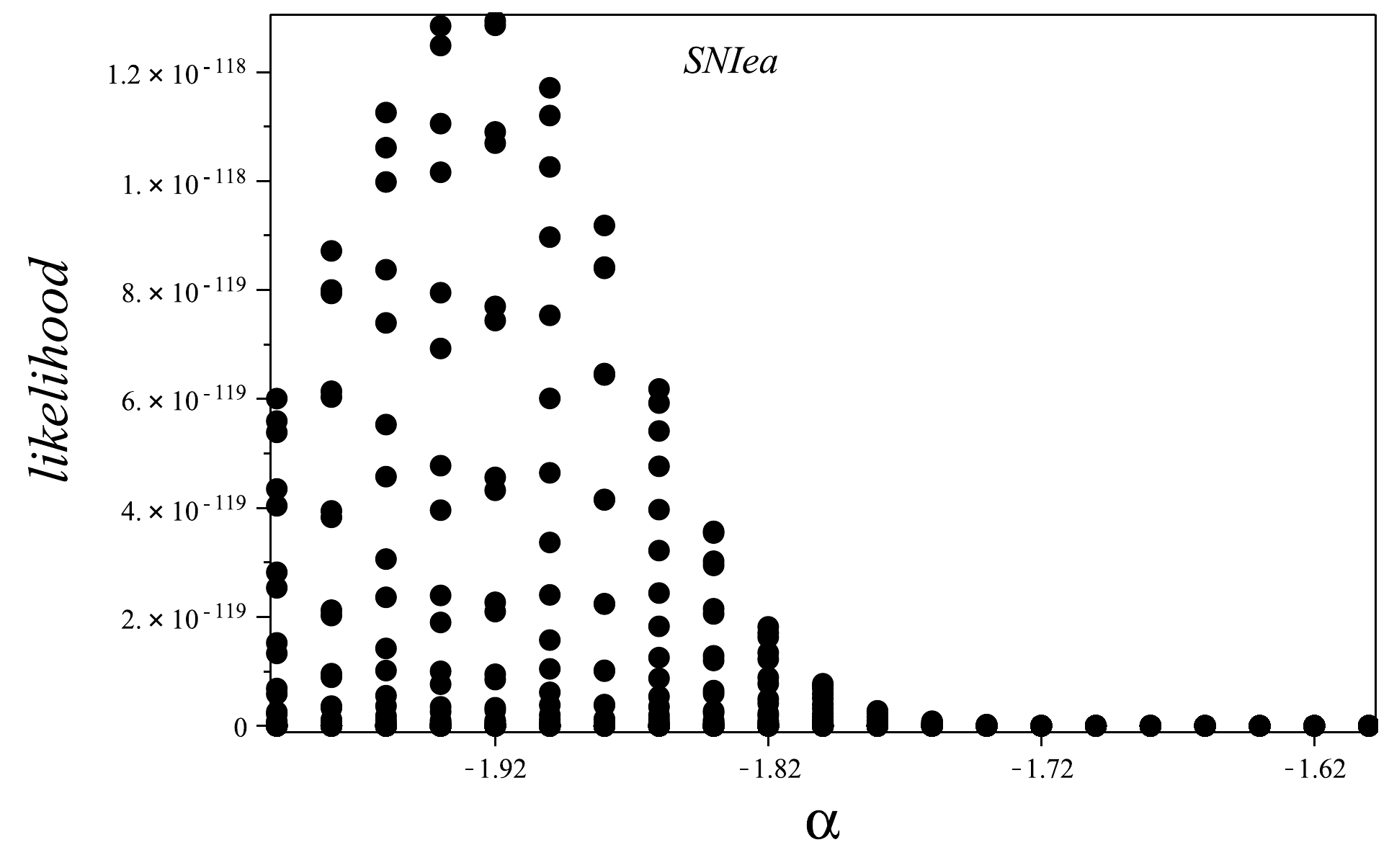}} \goodgap
\subfigure{\includegraphics[width=7cm]{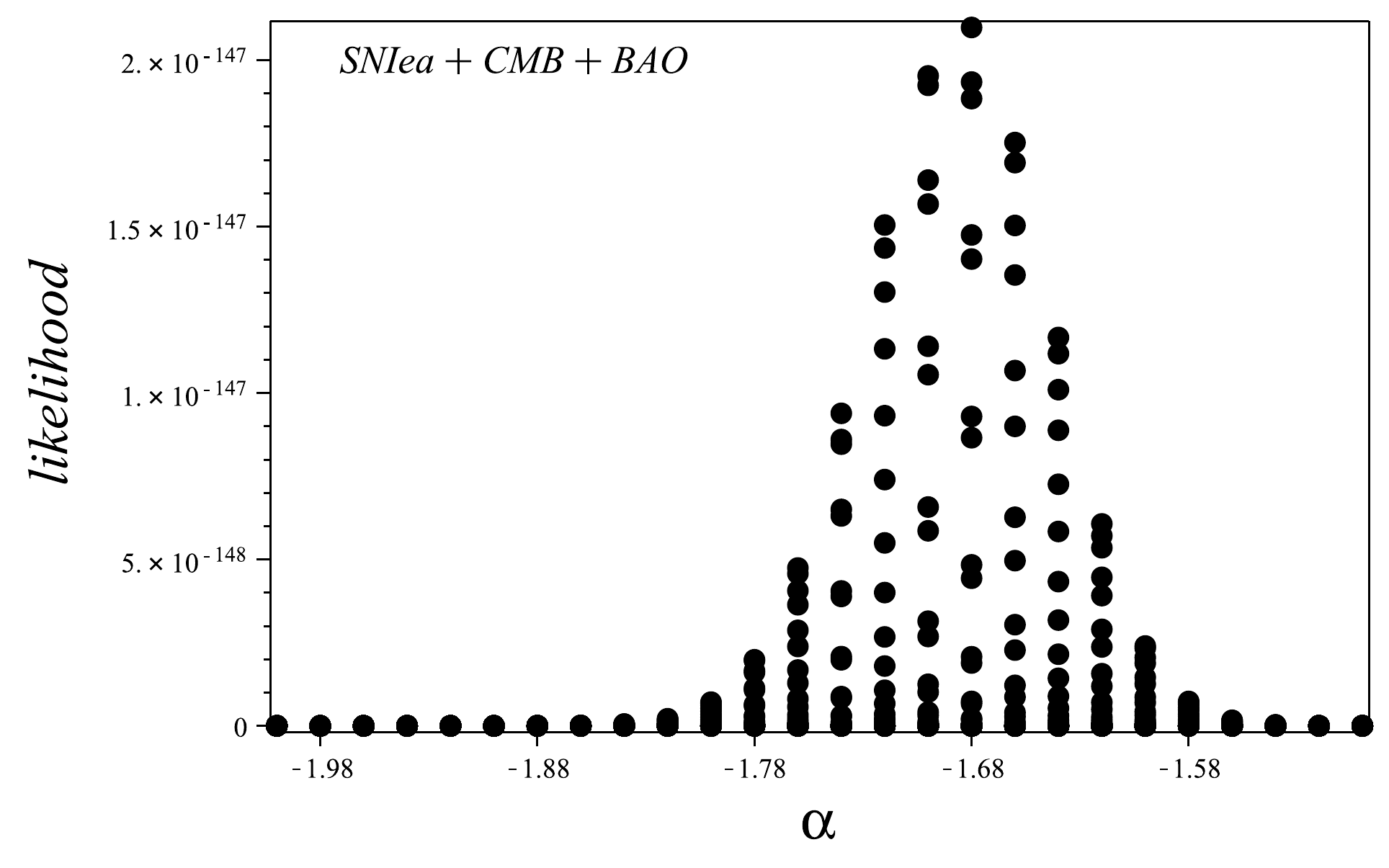}} \goodgap\\
\subfigure{\includegraphics[width=7cm]{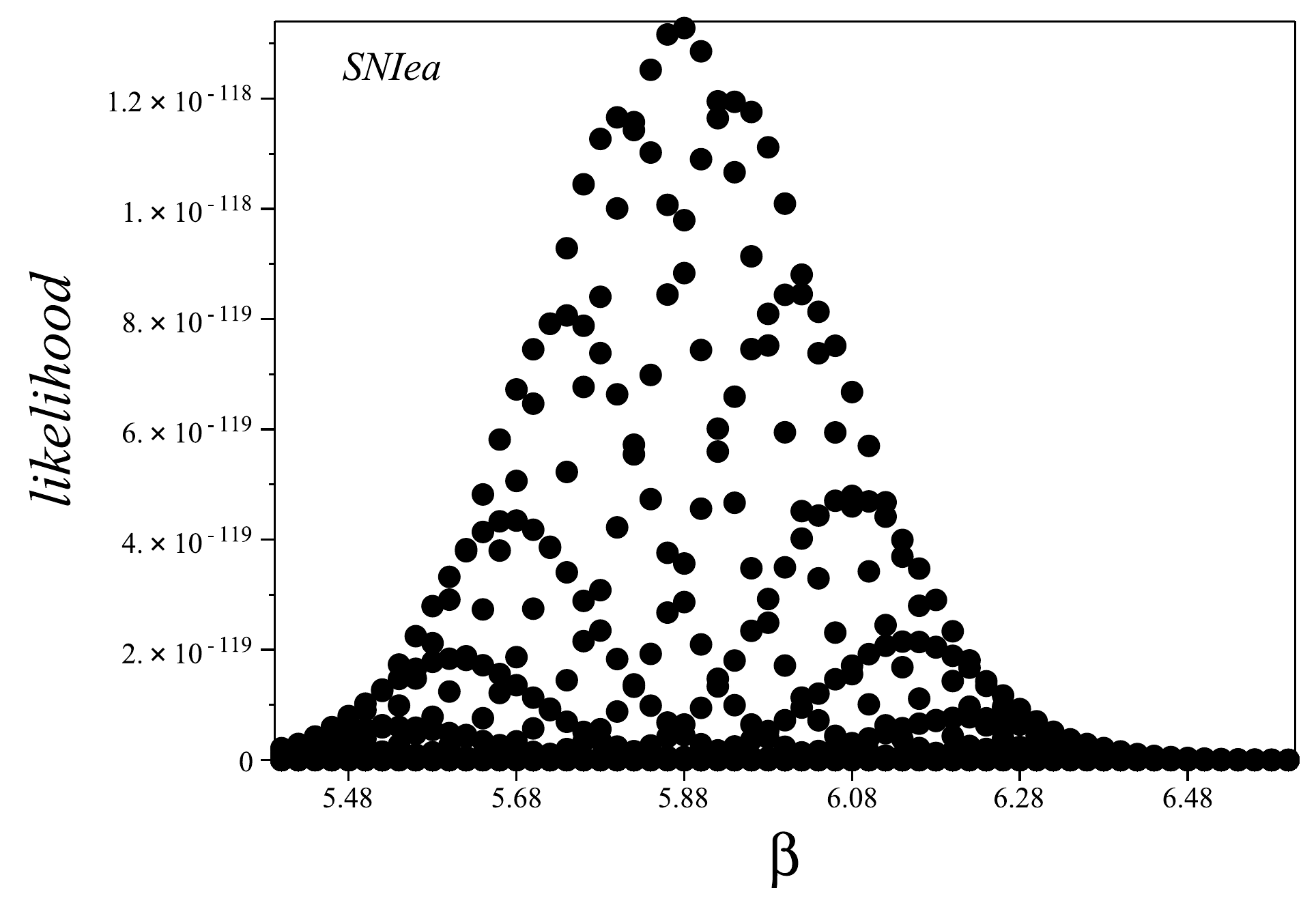}} \goodgap
\subfigure{\includegraphics[width=7cm]{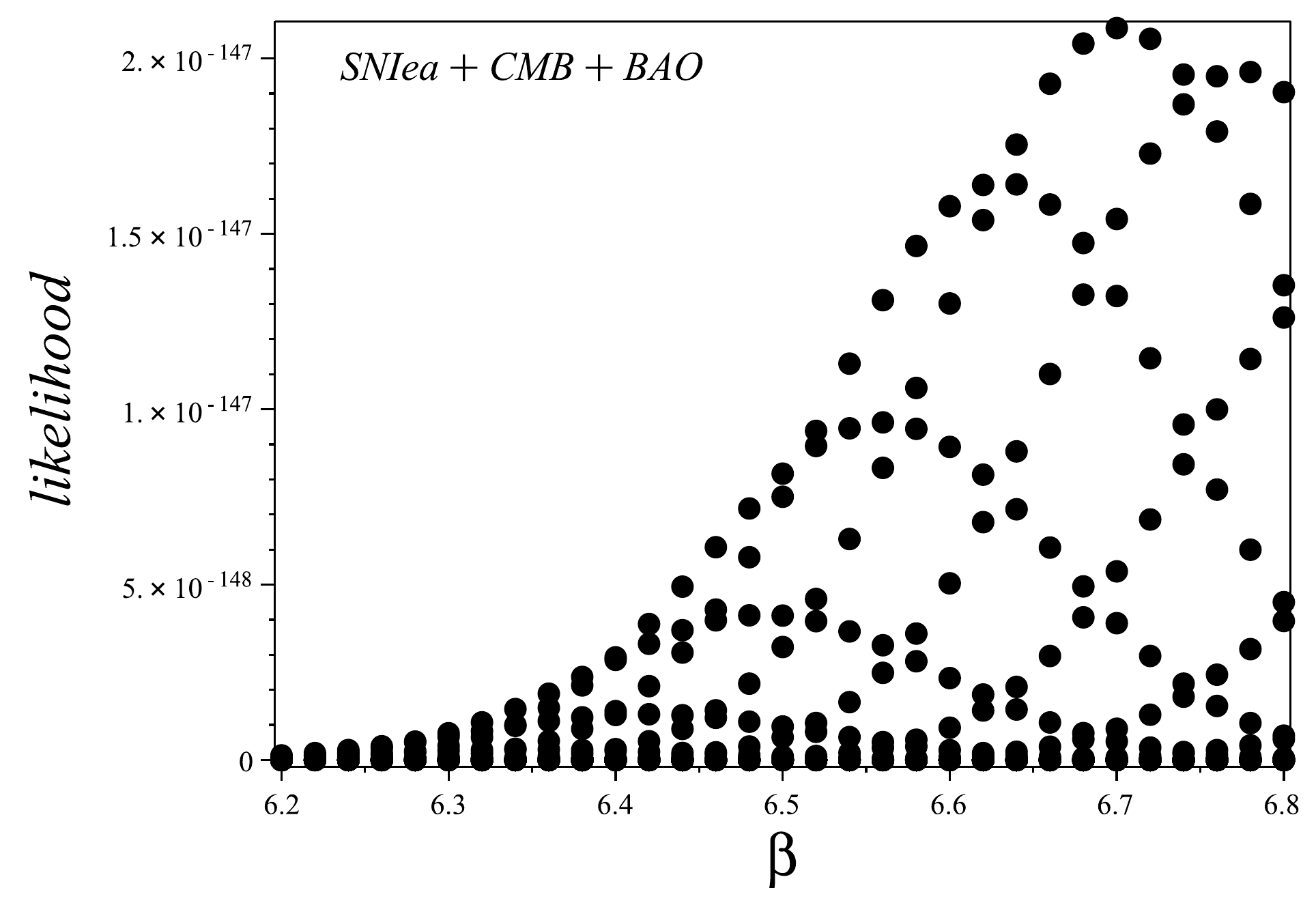}} \goodgap\\
\subfigure{\includegraphics[width=7cm]{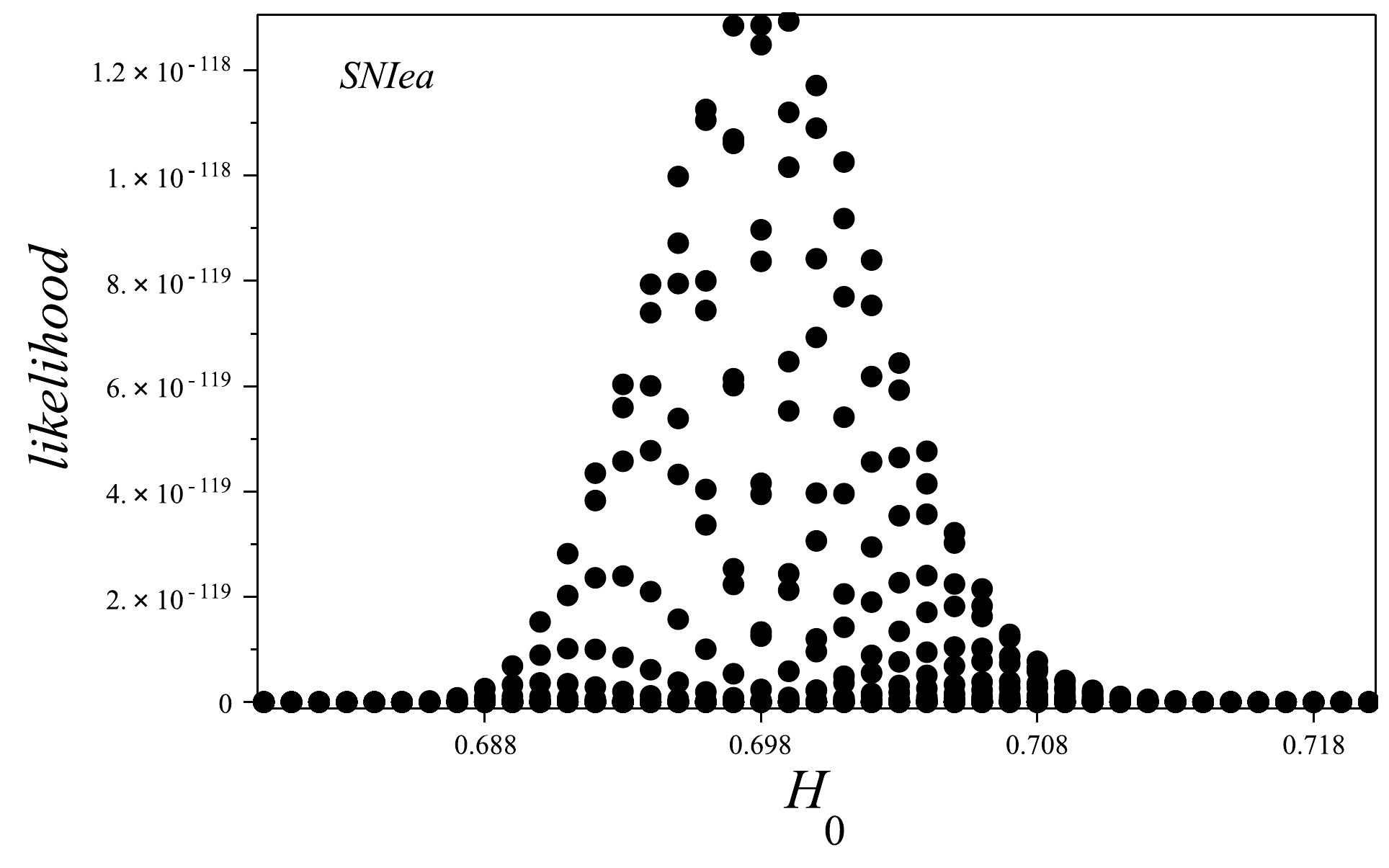}} \goodgap
\subfigure{\includegraphics[width=7cm]{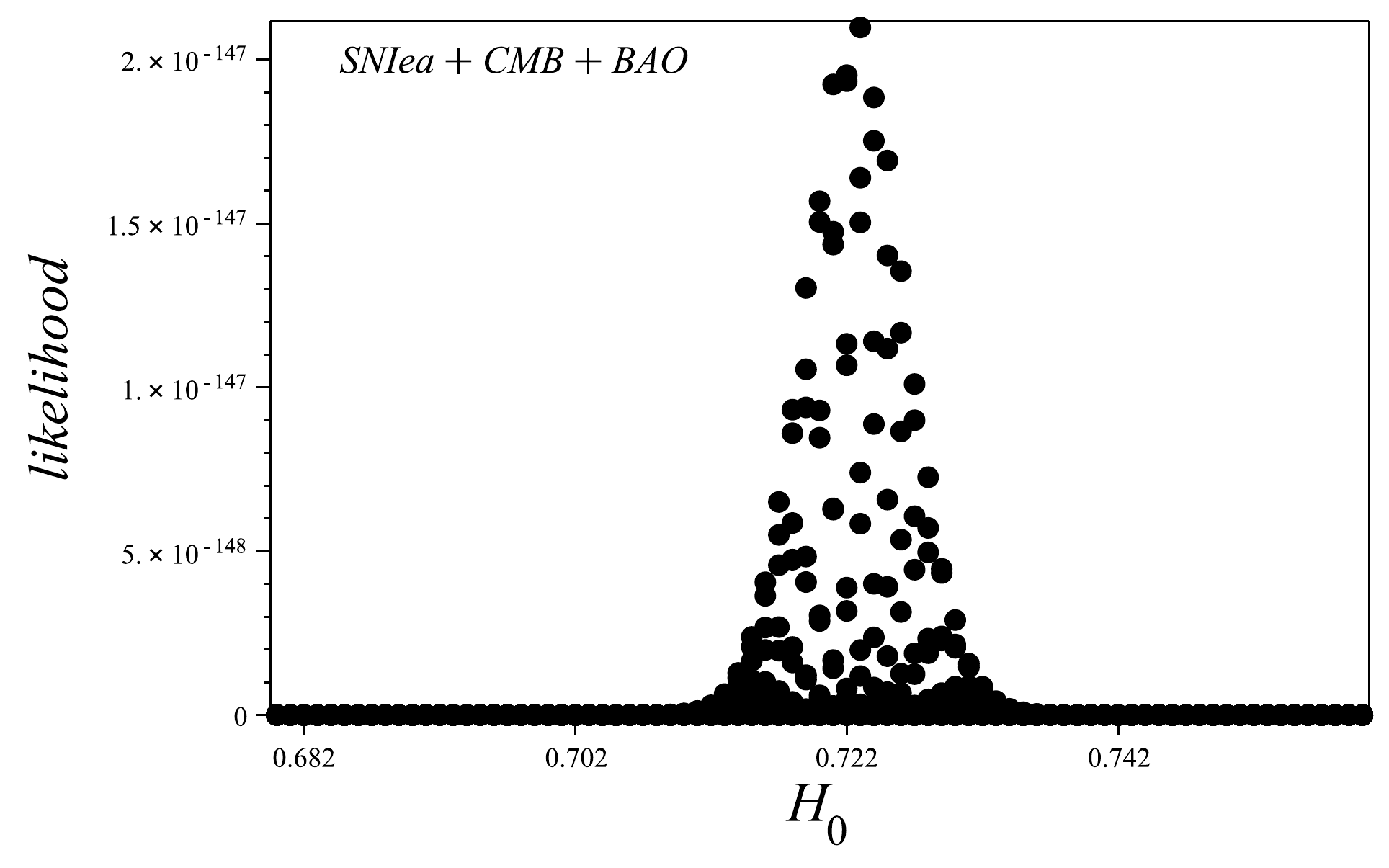}} \goodgap\\
\caption{The best-fitted two dimension likelihood and confidence level for $\alpha$, $\beta$ and $H_{0}$ for power law functions}
\label{fig: clplots}
\end{figure*}
\begin{figure*}
\centering
\subfigure{\includegraphics[width=7cm]{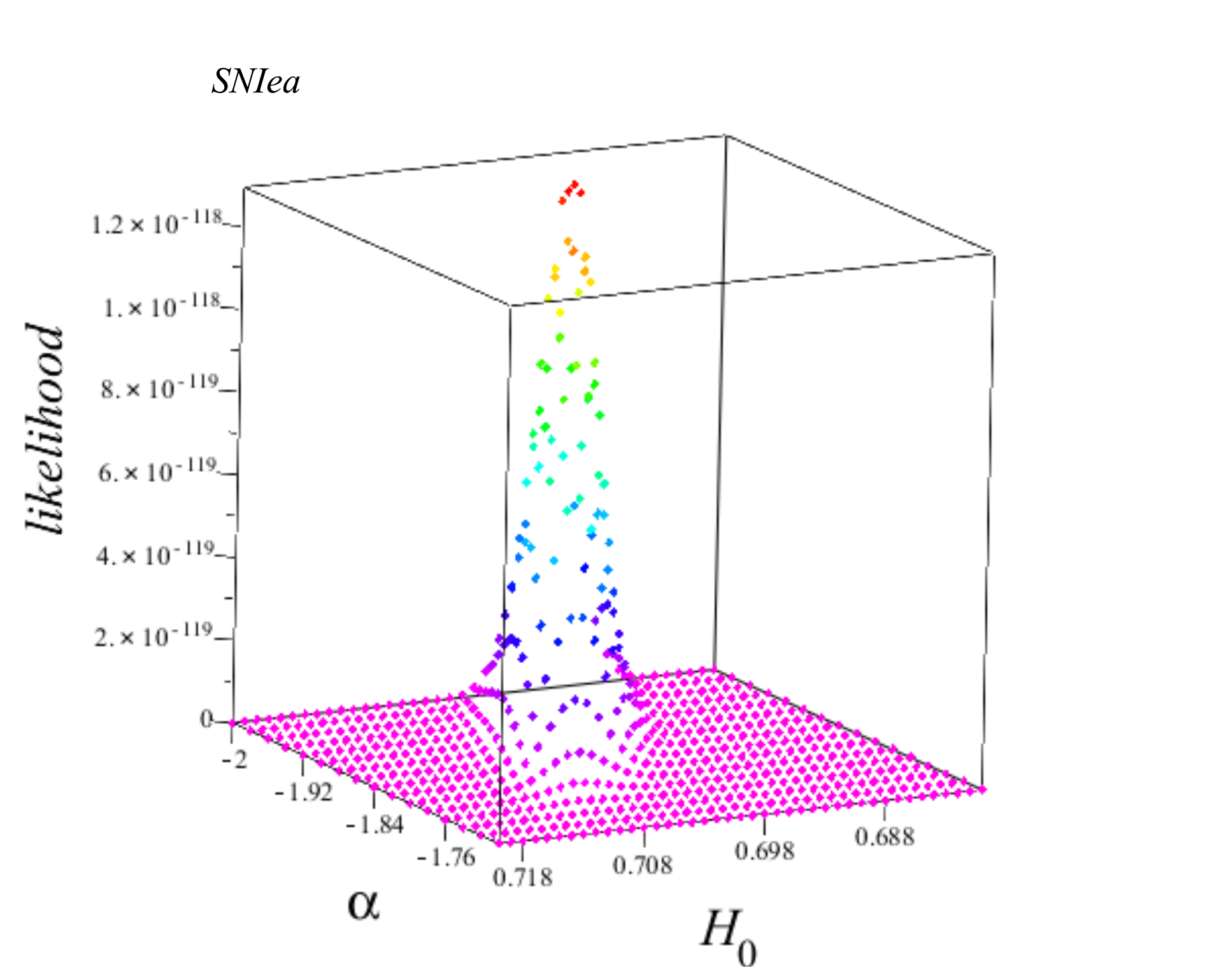}} \goodgap
%\subfigure{\includegraphics[width=7cm]{likeahexp.eps}} \goodgap\\
\subfigure{\includegraphics[width=7cm]{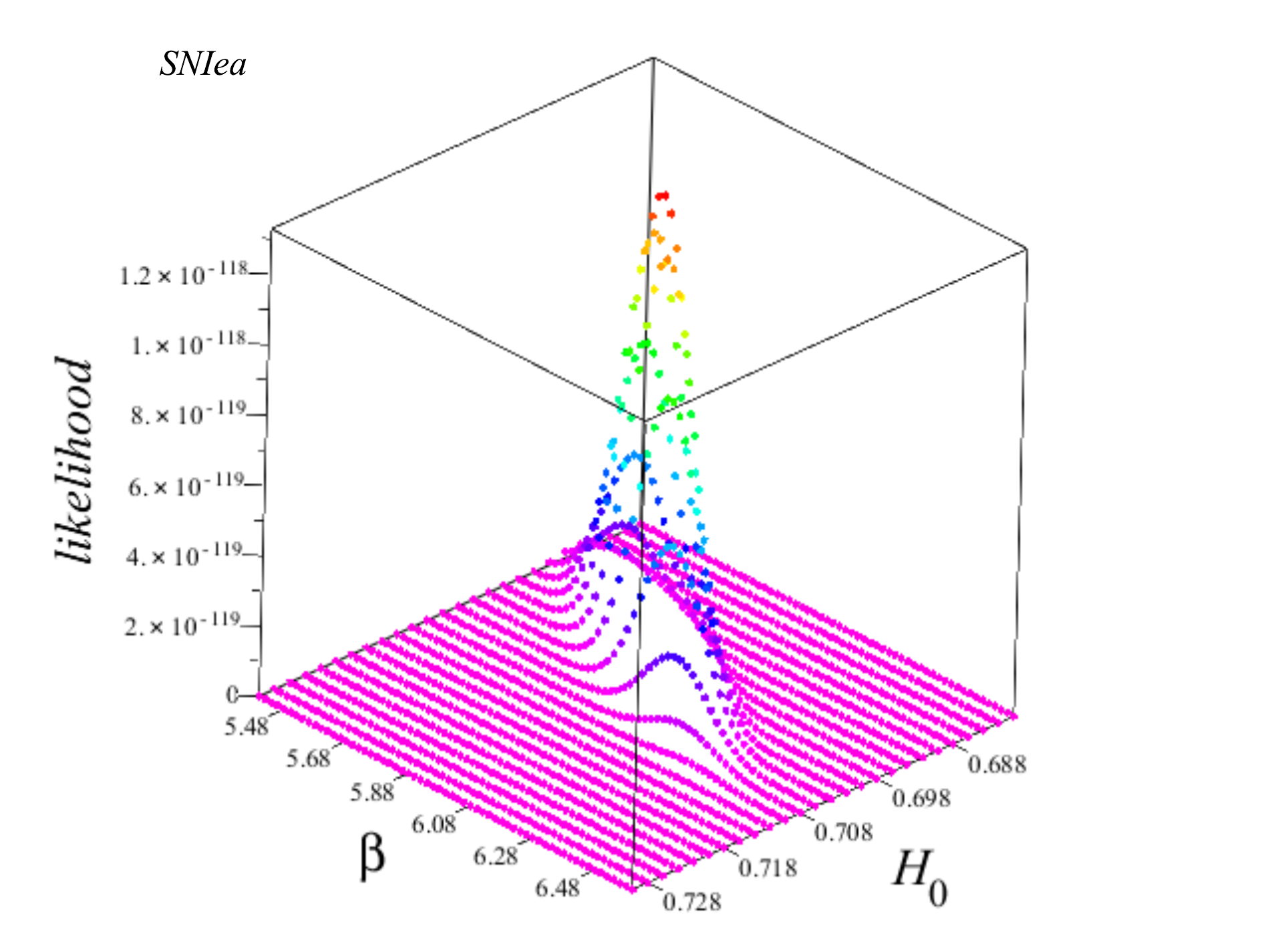}} \goodgap
\subfigure{\includegraphics[width=7cm]{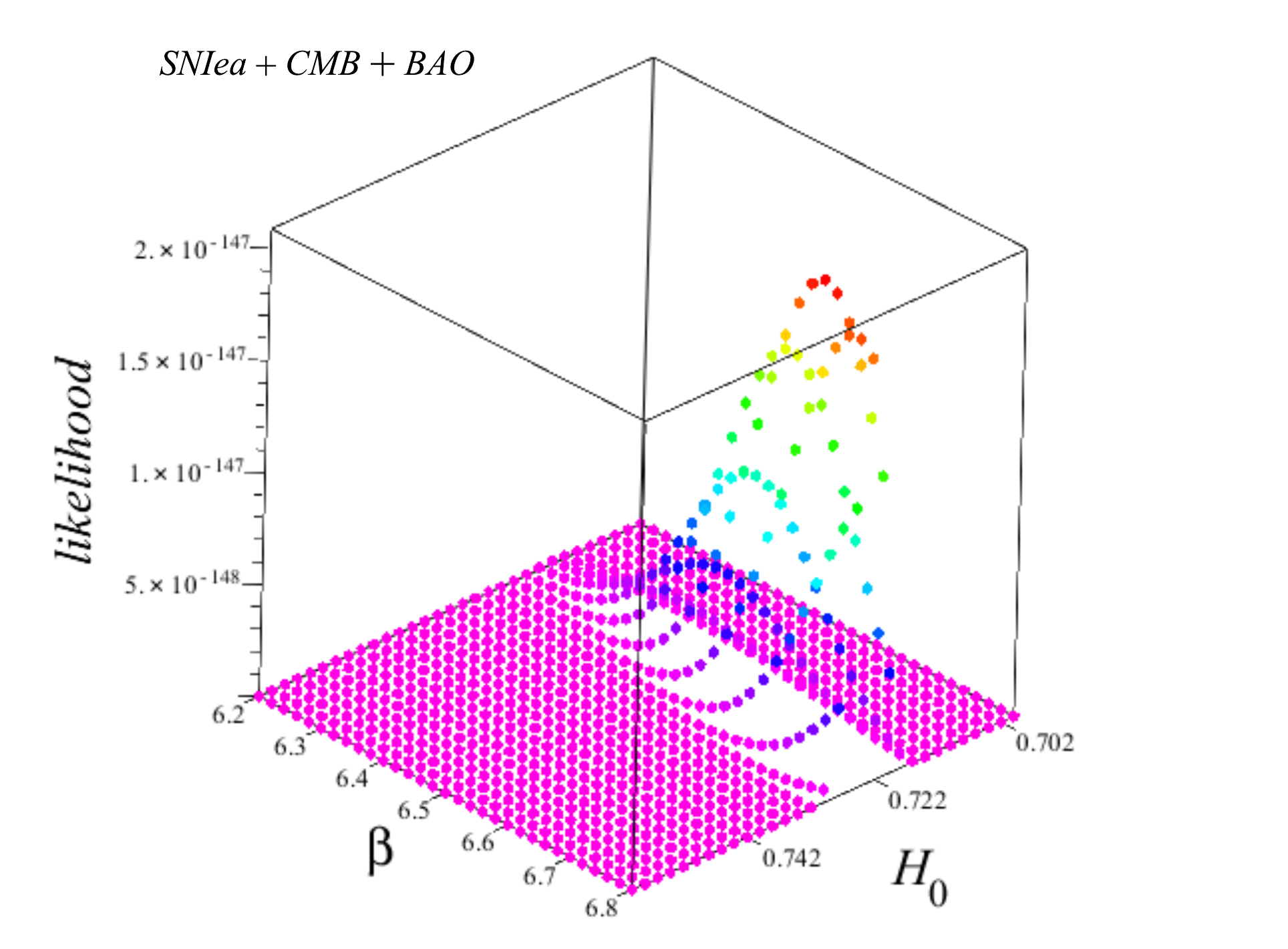}} \goodgap\\
\subfigure{\includegraphics[width=7cm]{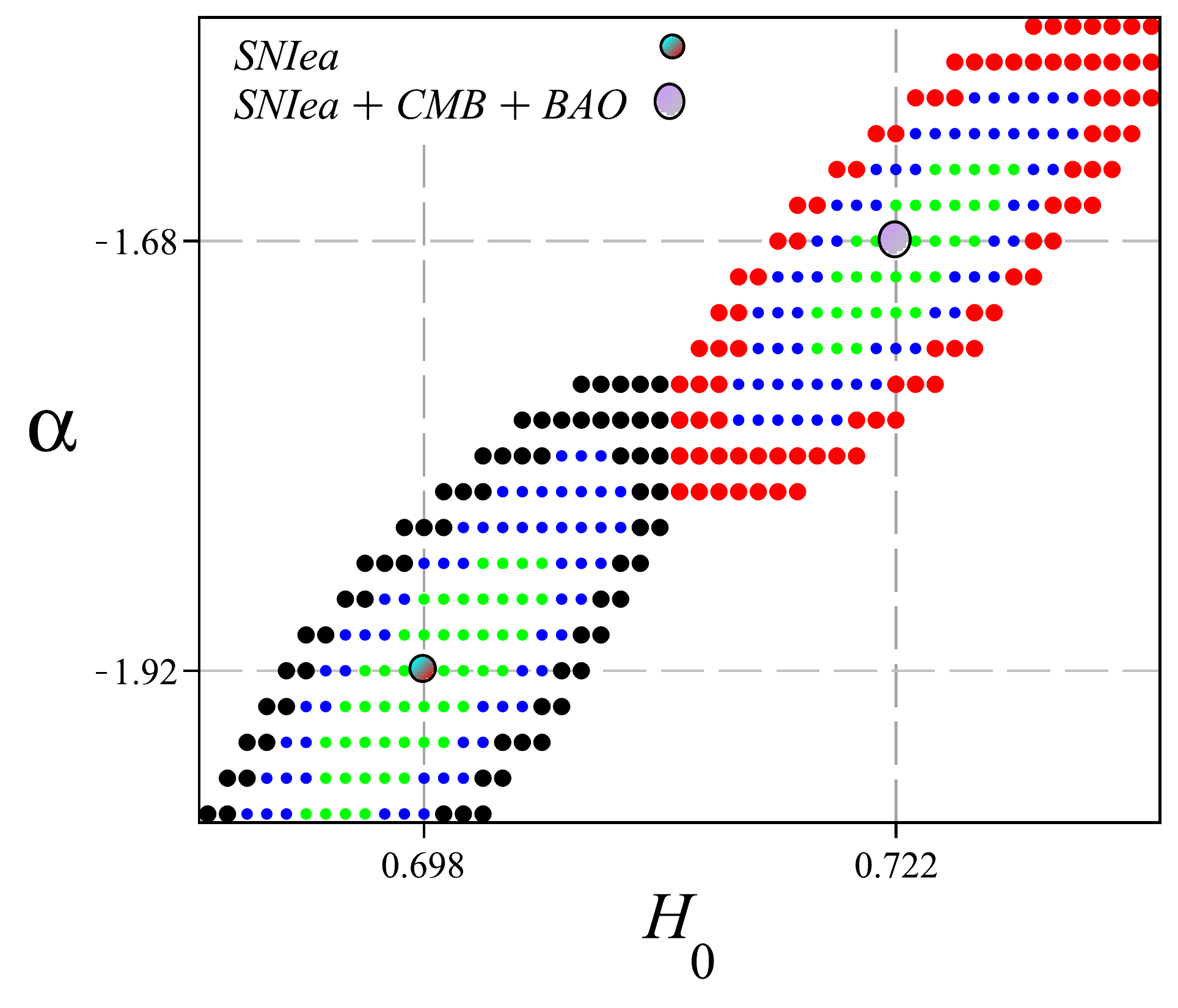}} \goodgap
\subfigure{\includegraphics[width=7cm]{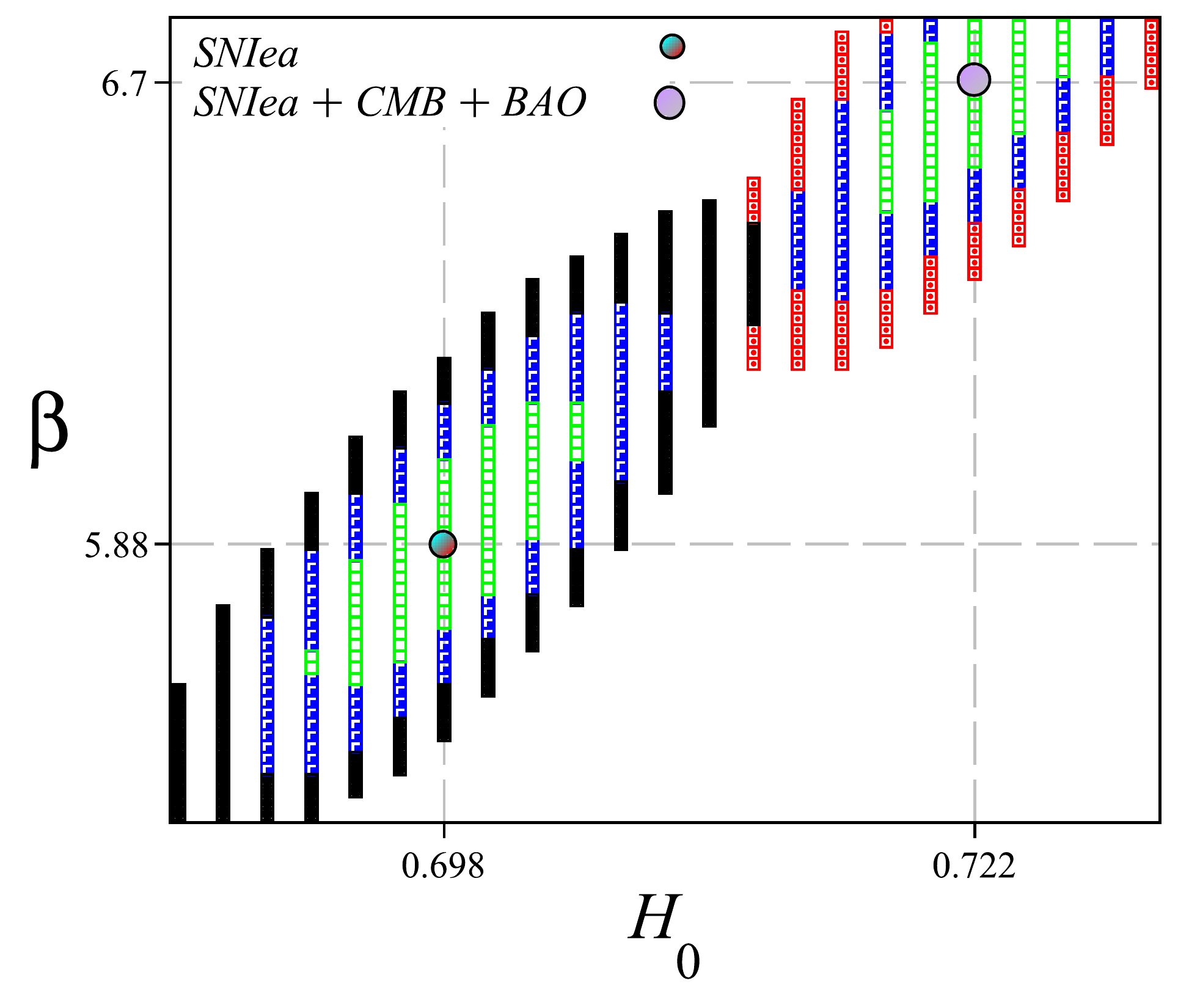}} \goodgap\\
\caption{The best-fitted two dimension likelihood and confidence level for $\alpha$, $\beta$ and $H_{0}$ for power law functions}
\label{fig: clplots}
\end{figure*}
In Fig. 5, the the numerically calculated distance modulus, $\mu(z)$, in the model is best fitted with the observational data for the model parameters $\alpha$, $\beta$ and initial conditions for $\Omega_F(0)$, $\Omega_{\dot{\phi}}(0)$ and $H(0)$ using $\chi^2$ method in both cases of power law and exponential functions.\\

\begin{figure*}
\centering
\subfigure{\includegraphics[width=7cm]{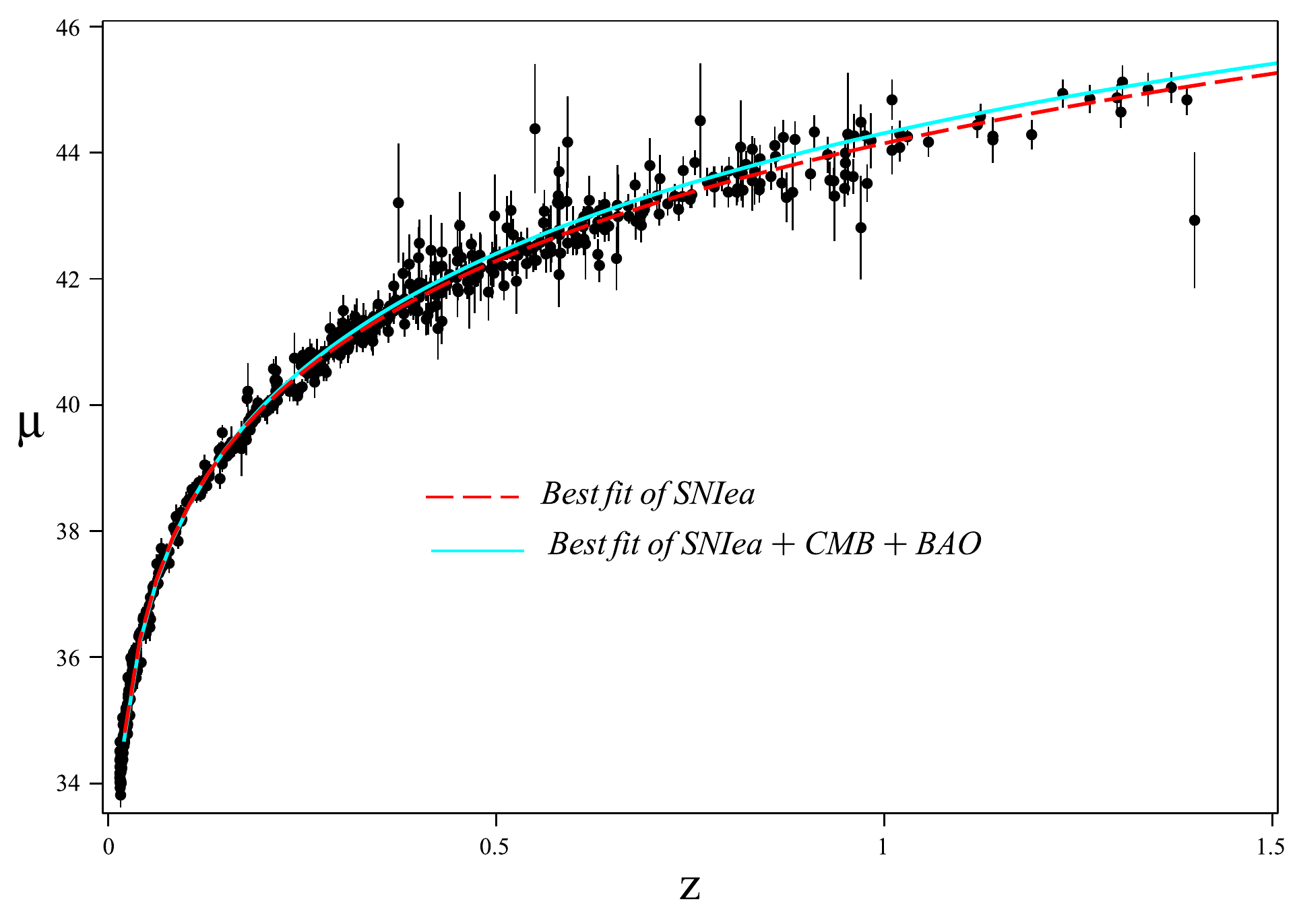}} \goodgap
\subfigure{\includegraphics[width=7cm]{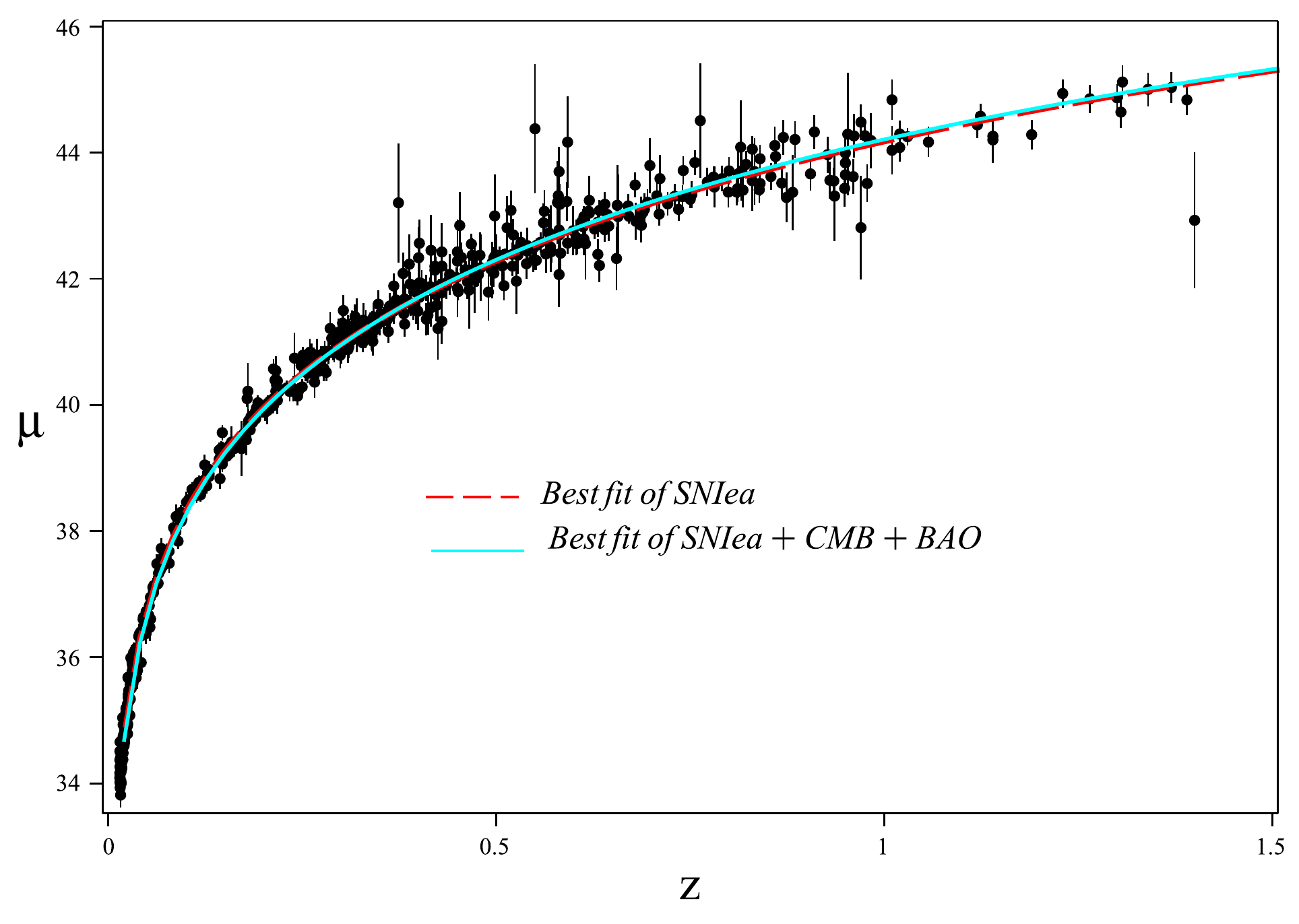}} \goodgap\\
\caption{The best-fitted distance modulus $\mu(z)$ plotted as function of redshift for $F(\phi)$ and $V$ left) exponential, right) power-law}
\label{fig: clplots}
\end{figure*}

Next, we examine the best fitted model against some other observational data. Unfortunately, the nature of dark matter and dark energy in the universe is such that direct observation are not possible. From current deceleration parameter of the universe we are able to determine the approximate value of the current EoS parameter. However to quantitatively predict the behavior of the EoS parameter of the universe in the past we need more observational data.

A direct verification of our model parameters is a comparison of the best fitted Hubble parameter obtained from numerical solution of the field equations with the observational data. In the following section, we test our model against the observational data for Hubble parameter and also the known dynamical behavior of the EoS parameter of the unvierse.

\section{Cosmological fitting results }

The effective EoS parameter in both cases of power law and exponential functions in terms of the new dynamical variables respectively are given by
\begin{eqnarray}\label{eos}
\omega_{eff}&=&-1+\frac{4\Omega_{\dot{\phi}}}{3(4\Omega_{\dot{\phi}}+\Omega_{F}^{2})}(3+\Omega_{F}
+\frac{\alpha-1}{\alpha}\Omega_{F}^{2}+3\Omega_{\dot{\phi}}+3 \Omega_{V}+\frac{\Omega_{F}^{2}}{\Omega_{\dot{\phi}}}-\frac{\beta \Omega_{V}\Omega_{F}^{2}}{2\alpha \Omega_{\dot{\phi}}}),\\
\omega_{eff}&=&-1+\frac{4\Omega_{\dot{\phi}}}{3(4\Omega_{\dot{\phi}}+\Omega_{F}^{2})}(3+\Omega_{F}
+\Omega_{F}^{2}+3\Omega_{\dot{\phi}}+3 \Omega_{V}+\frac{\Omega_{F}^{2}}{\Omega_{\dot{\phi}}}-\frac{\beta \Omega_{V}\Omega_{F}^{2}}{2\alpha \Omega_{\dot{\phi}}}).
\end{eqnarray}

With the best-fitted model parameters and initial conditions with the observational data in both cases, the effective EoS parameters are shown in Fig. 6.

\begin{figure*}
\centering
\subfigure{\includegraphics[width=7cm]{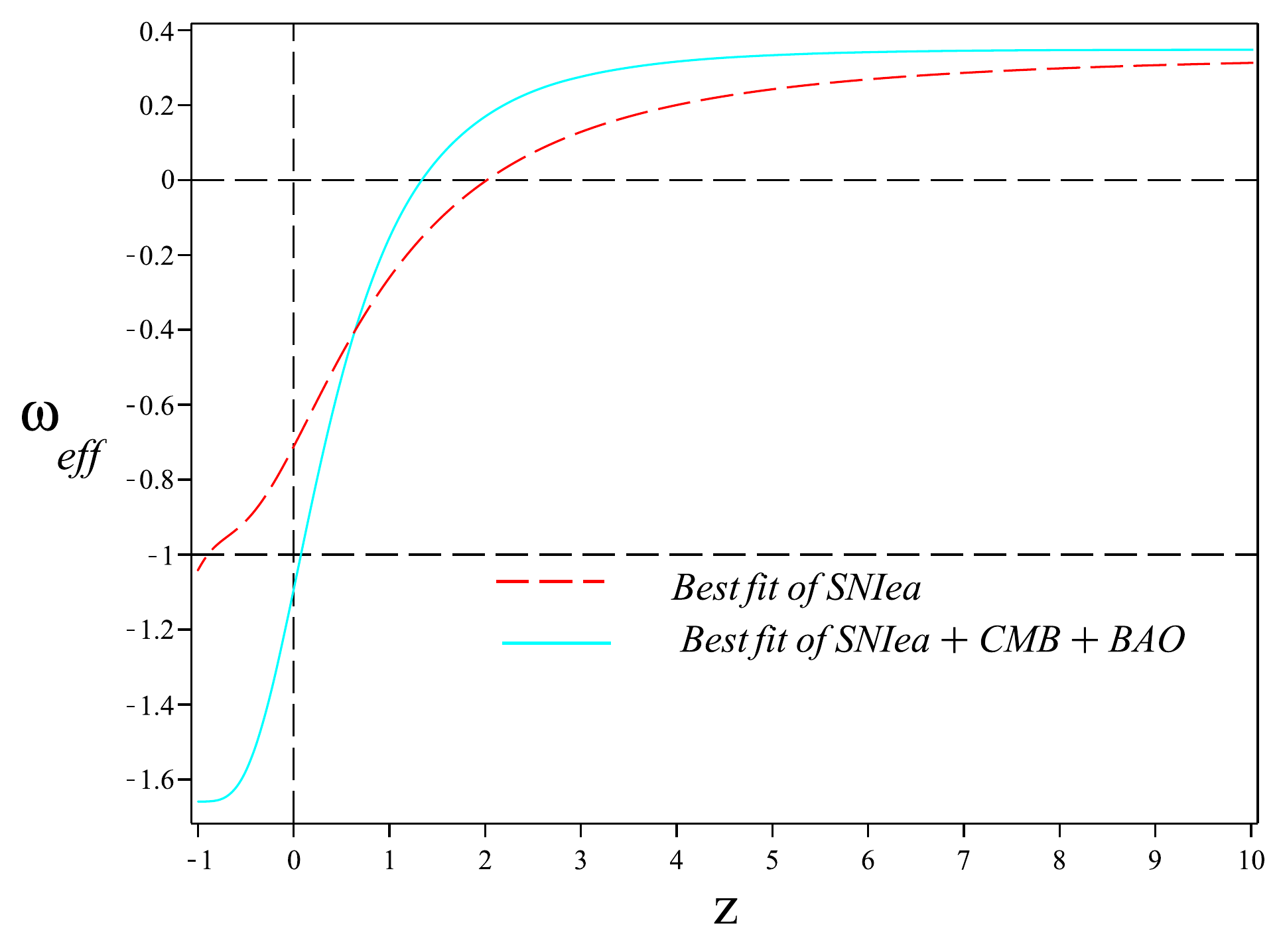}} \goodgap
\subfigure{\includegraphics[width=7cm]{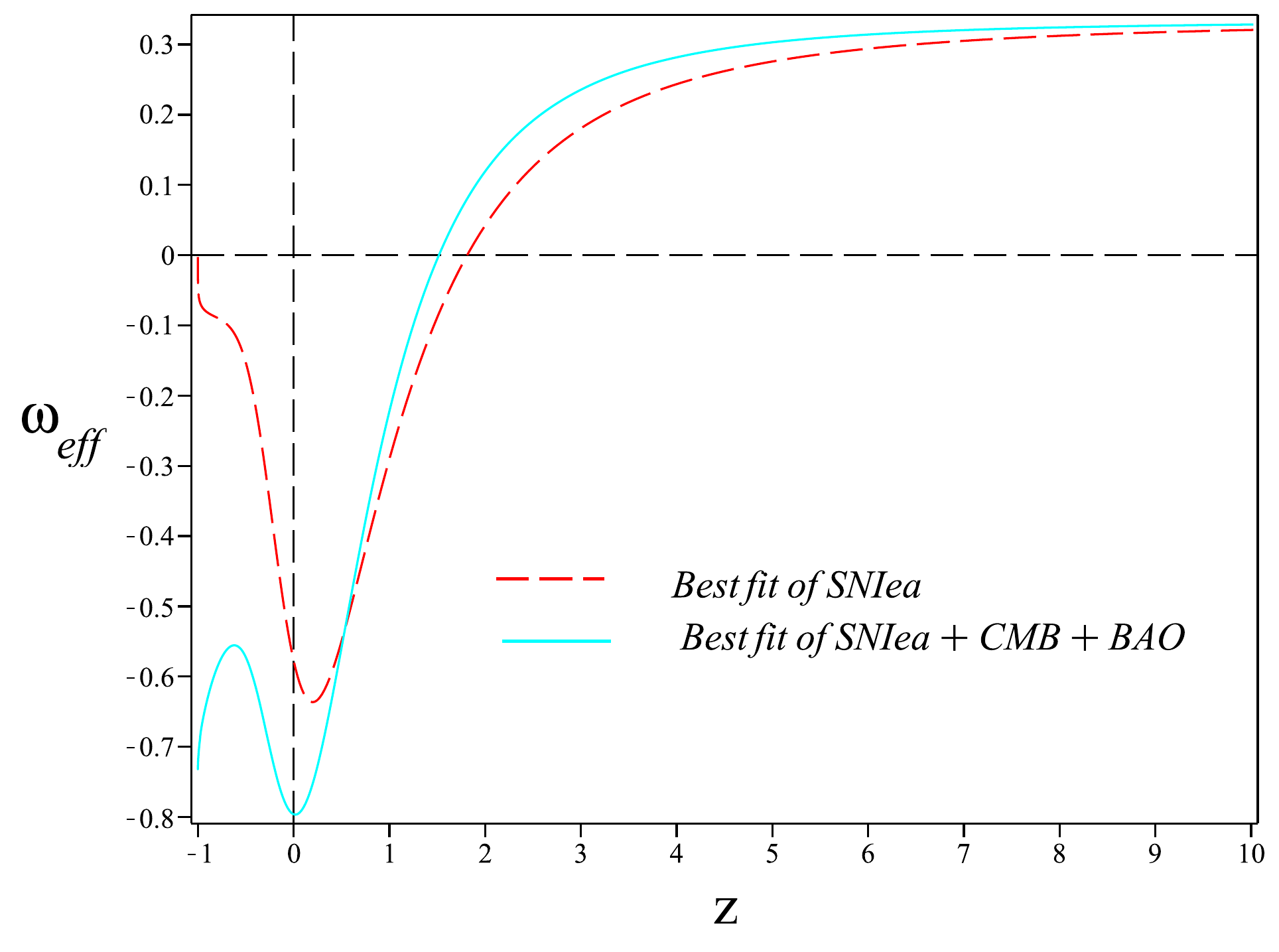}} \goodgap\\
\caption{The best-fitted effective $EoS$ parameter plotted as function of redshift for $F(\phi)$ and $V$ left) exponential, right) power-law}
\label{fig: clplots}
\end{figure*}

As can be seen, a common feature in both cases of power law and exponential is that the best fitted EoS parameters in high redshifts show radiation dominated era. However, in power-law case, phantom crossing never occurs in the past or future whereas in exponential case, the best fitted parameter with Sne Ia exhibits phantom crossing in the future and the best fitted EoS parameter with Sne Ia+ CMB + Bao shows phantom crossing in the past.

A second cosmological test with the observational data for Hubble parameter is shown in Fig. 7 in both power paw and exponential cases. From the graphs we observe that in both cases, our model is in good agreement with the observational data.

\begin{figure*}
\centering
\subfigure{\includegraphics[width=7cm]{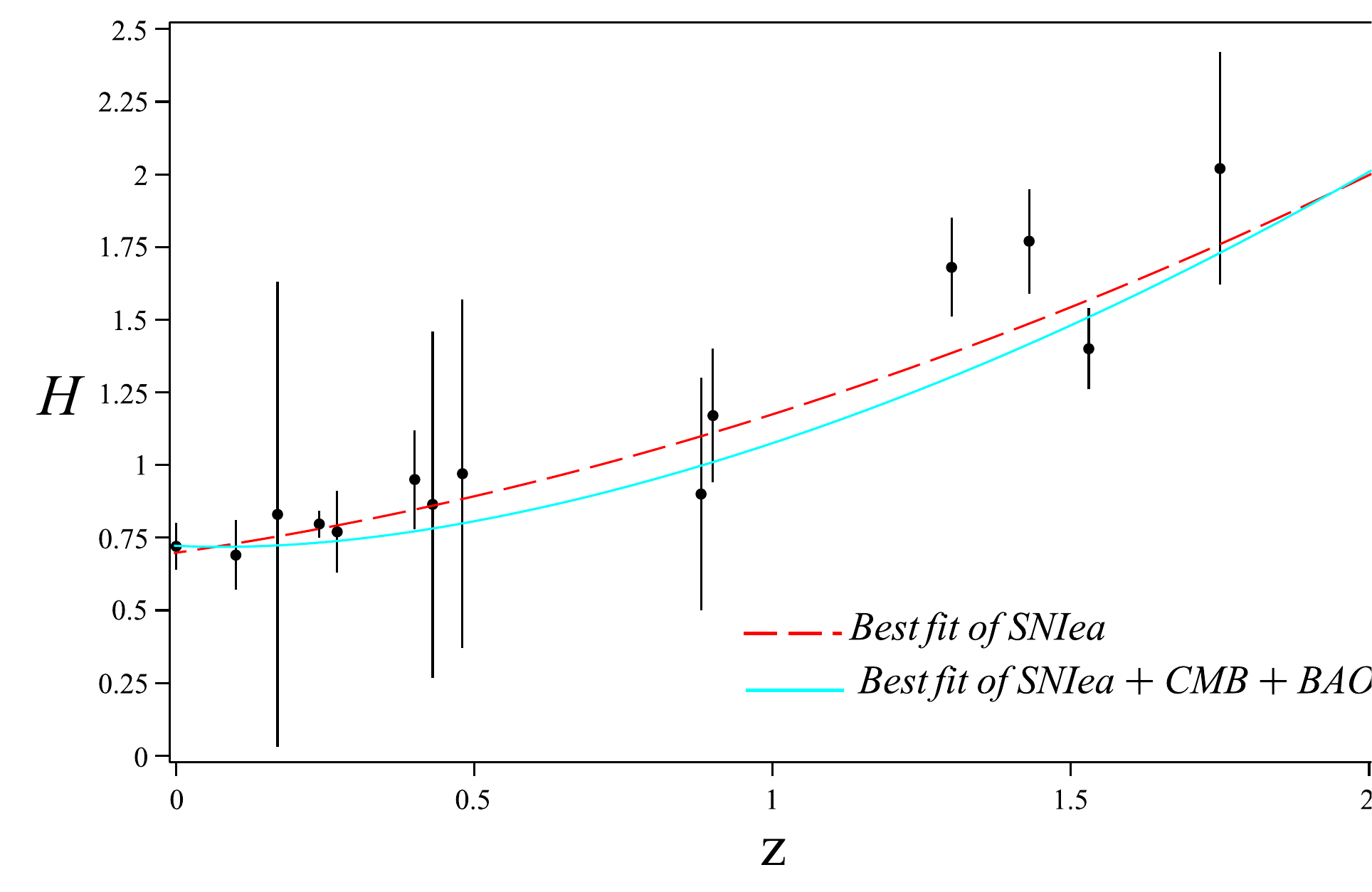}} \goodgap
\subfigure{\includegraphics[width=7cm]{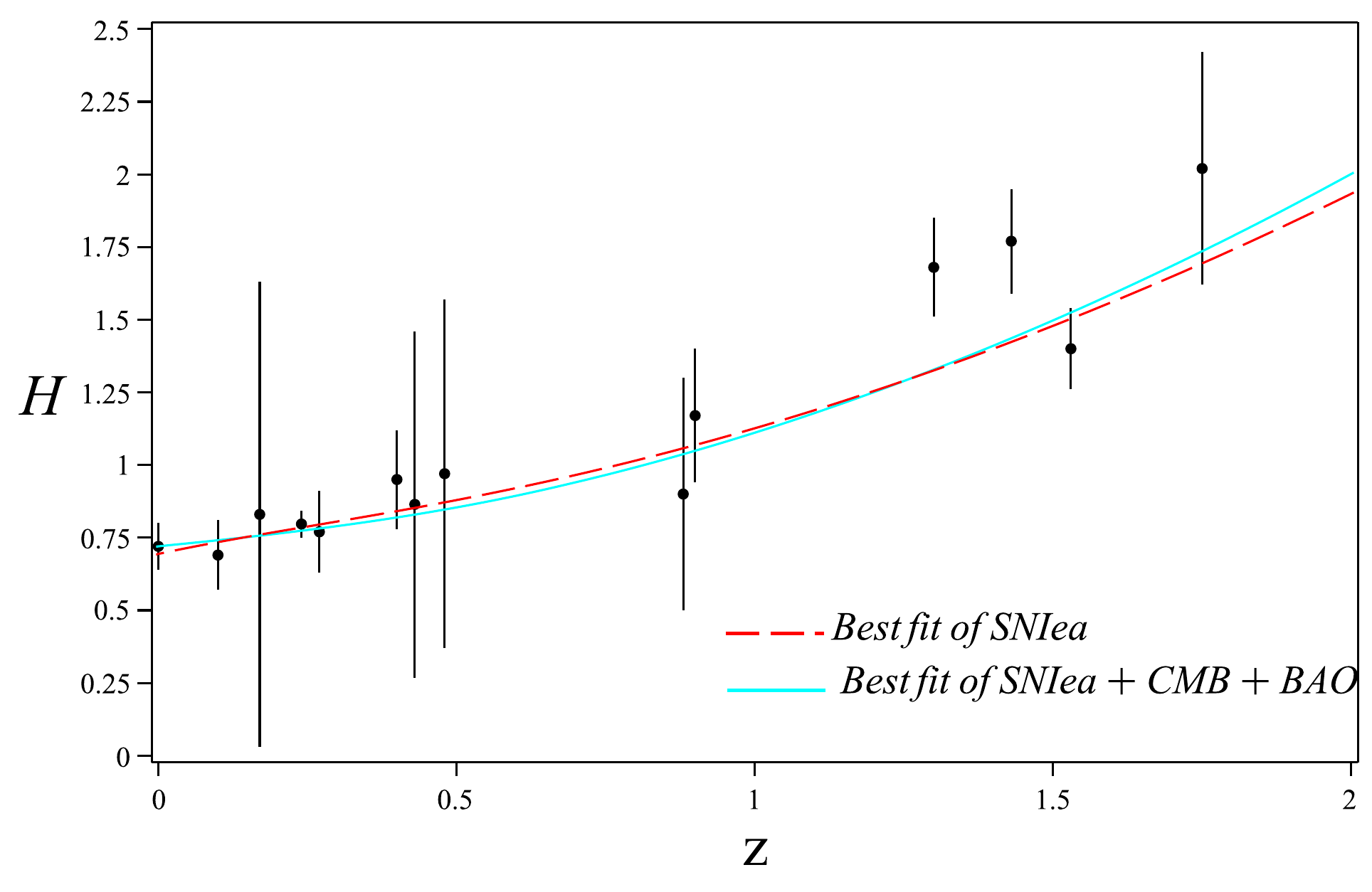}} \goodgap\\
\caption{The best-fitted  $H(z)$ plotted as function of redshift for $F(\phi)$ and $V$ left) exponential, right) power-law}
\label{fig: clplots}
\end{figure*}

In addition, the behavior of the best fitted reconstructed functions $F(\phi)$ and potential $V(\phi)$ versus $\phi$ in both exponential and power law cases are also shown in Fig.8.

\begin{figure*}
\centering
\subfigure{\includegraphics[width=7cm]{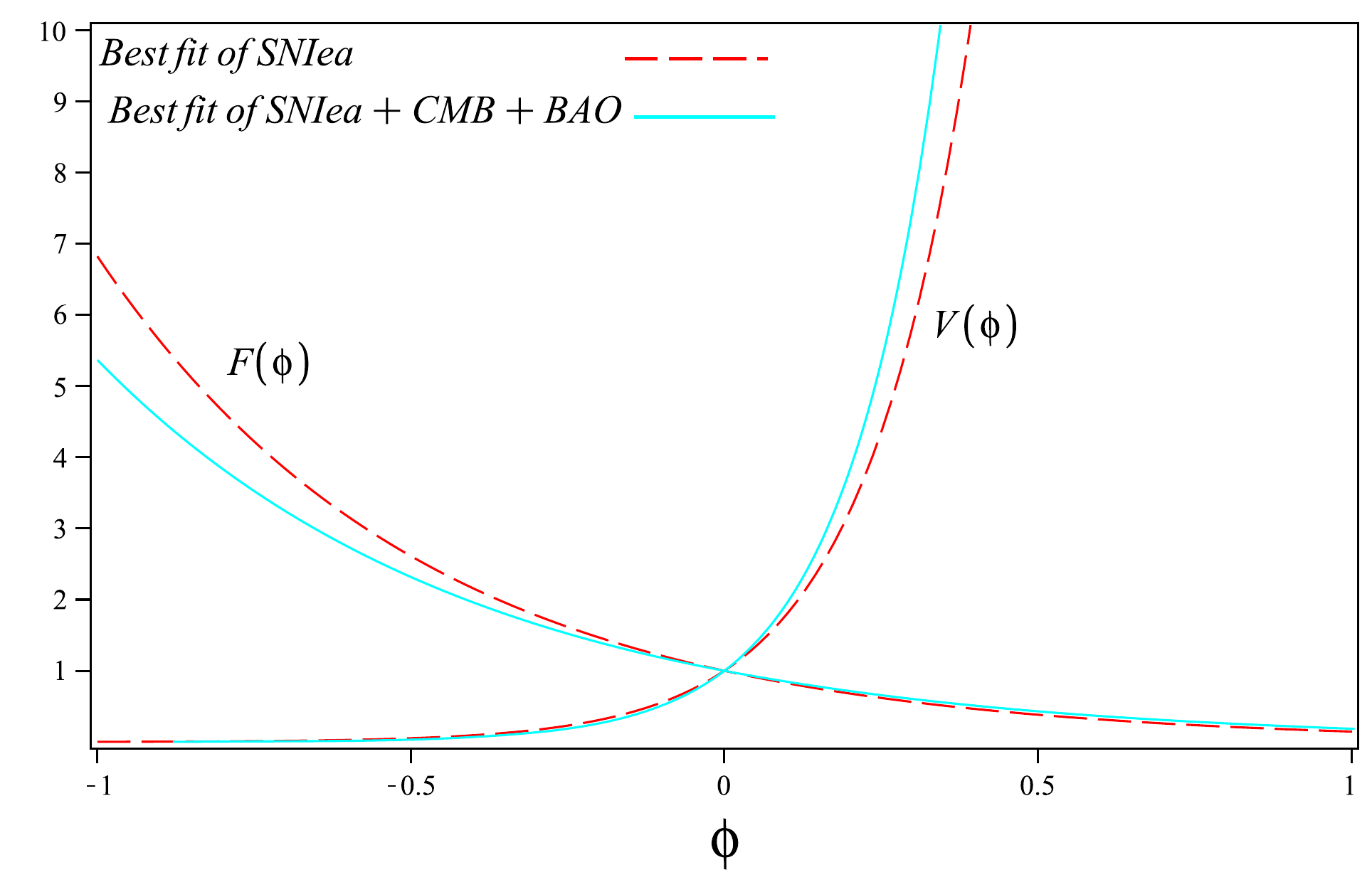}} \goodgap
\subfigure{\includegraphics[width=7cm]{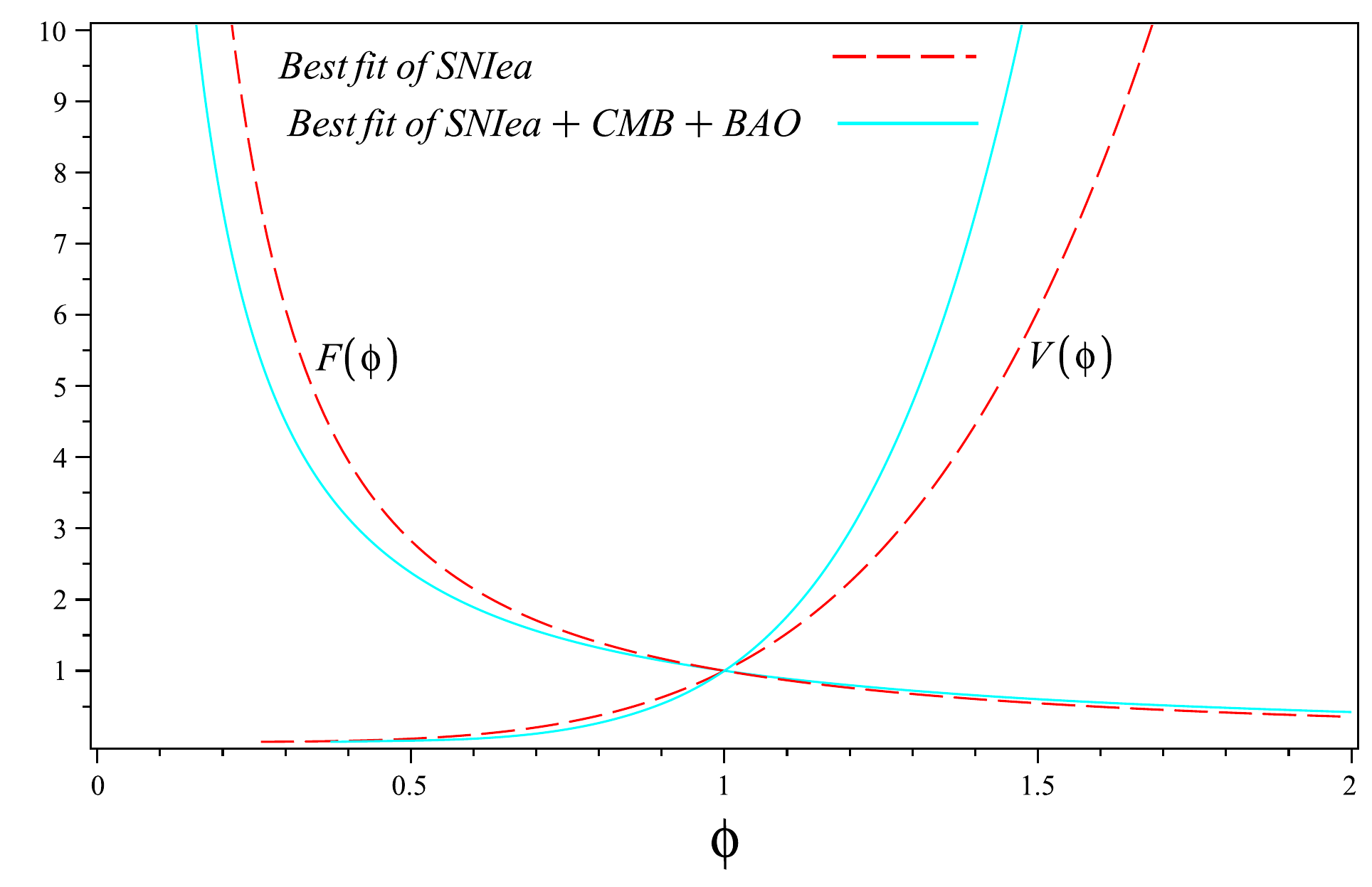}} \goodgap\\
\caption{The best-fitted reconstructed $F(\phi)$ and $V(\phi)$ plotted as function of redshift for left) exponential forms, right) power-law forms}
\label{fig: clplots}
\end{figure*}

\section{Discussion and summary}

This paper is to present the dynamics of the scalar–-tensor gravity theories in cosmology. We assume two different scenarios for the scalar field function $F(\phi)$ coupled to the geometry and the potential $V(\phi)$; in the exponential and power law forms. In both cases, by representing the model in terms of the new dynamical variables we simultaneously solve the field equations in flat FRW cosmology and best fit the model parameters and also the initial conditions with the observational data for distance modulus using the $\chi^2$ test. Best fitting the model parameters provides physically reliable and observationally verified solutions. The best fitted effective EoS parameters, $\omega_{eff}$, in both scenarios exhibit a late accelerating universe.

In power law case, the universe begins from a decelerating state in high redshift (radiation dominated era), transits to an accelerating state at about $z\simeq 1$ in the past (quintessence phase with $\omega_{eff}>-1$) and again in near future re--enter the deceleration state. In this scenario, the universe never experience phantom crossing and phantom phase. Similarly, in the case of exponential function parameterizations, the universe begins from deceleration state in the high redshift and enter the acceleration state in the near past at about the same redshift. However, it will also transit to the phantom phase ($\omega_{eff}<-1$) in the future for the best fitted effective EoS parameter with Sne Ia data and in the past for the best fitted with Sne Ia+ CMB + Bao data. However,  in power law case, constrained by only Sne Ia, the EoS parameter in future transits from phantom to quintessence  phase and finally to matter dominated phase. From the graph we also observe that the current values for effective EoS parameter in both exponential and power law cases are between $-0.4$ and $-1.1$.

The phantom crossing feature is impossible to reproduce in the context of single field
quintessence or even phantom models. It is interesting that in our scalar--tensor model with phantom field (negative kinetic term), and best fitted power law $F(\phi)$, the crossing occurs while the requirement of $F(\phi)>0$ is guaranteed. In this paper, we have also constructed experimentally viable function $F(\phi)$ and  potential $V(\phi)$ with the best fit parameterizations obtained from the recent SNe Ia, CMB and Bao dataset. . From Fig. 7, the model shows a good agreement with the observational data for Hubble parameter. Fig. 8 shows the behaviour of the the best fitted $F(\phi)$ and $V(\phi)$ for both exponential and power. The best fitted reconstructed functions show monotonic increasing/decreasing behavior in each model with respect to the scalar field $\phi$. The positive $F(\phi)$ in both cases is in agreement with the theoretical requirement.


\begin{thebibliography}{99}


\bibitem{Reiss} A.G. Reiss et al, Astron. J. 116, 1009 (1998) ; S. Perlmutter et al, Astrophys
J. 517 565(1999); J. L. Tonry et al, Astrophys J. 594, 1-24 (2003)

\bibitem{Bennet} C. I. Bennet et al, Astrophys J. Suppl. 148:1, (2003); C. B.
 Netterfield et al, Astrophys J. 571, 604 (2002); N. W. Halverson et
al, Astrophys J. 568, 38 (2002)

\bibitem{Pope} A. C. Pope, et. al, Astrophys J. 607 655 (2004)

\bibitem{Riess2} A.G. Riess et al., Astrophys. J. 607 (2004) 665; R. A. Knop et al,
Astrophys. J. 598 102 (2003)

\bibitem{Abazajian}K. Abazajian et al, Astron. J. 129 1755 (2005); Astron. J. 128
502(2004); Astron. J. 126 (2003) 2081; M. Tegmark et al, Astrophys. J. 606 702 (2004)

\bibitem{Allen}S.W. Allen, R. W. Schmidt, H. Ebeling, A. C. Fabian and L. van
Speybroeck, Mon. Not. Roy. Astron. Soc. 353 457 (2004)

\bibitem{Bennett} C.L. Bennett et al, Astrophys. J. Suppl. 148 1(2003)

\bibitem{Spergel} D. N. Spergel, et. al., Astrophys J. Supp. 148 175 (2003)


\bibitem{Seljak} U. Seljak et al, Phys. Rev. D 71 (2005) 103515; M. Tegmark, JCAP
0504 001 (2005)

\bibitem{Setare} M. R. Setare, Phys. Lett. B644:99-103 (2007); J. Sadeghi, M. R. Setare, A. Banijamali,
Eur. Phys. J. C64:433-438 (2009); J. Sadeghi, M. R. Setare,  A. Banijamali,  Phys. Lett. B678:164-167 (2009)

\bibitem{Fujii} Y. Fujii, Prog. Theor. Phys.118:983-1018 (2007)


\bibitem{capelo}S. Capozziello, G. Lambiase, Grav.Cosmol. 6, 173-180 (2000)

\bibitem{Sahoo}B.K. Sahoo and L.P. Singh, Mod. Phys. Lett. A 17 (2002)
; B.K. Sahoo and L.P. Singh, Mod. Phys. Lett. A 18 2725(2003)
; B.K. Sahoo and L.P. Singh, Mod. Phys. Lett. A 19 1745 (2004)


\bibitem{Capozziello}
S. Capozziello, S. Carloni and A. Troisi, Recent Res. Dev. Astron. Astrophys. 1:625 (2003); S. Nojiri and S.D. Odintsov, Phys. Rev. D 68 123512(2003) ; Phys. Lett. B 576 5 (2003)

\bibitem{Faraoni} V. Faraoni, Phys. Rev. D 75 067302 (2007); J.C.C. de Souza, V.
Faraoni, Class. Quant. Grav. 24 3637 (2007) ; A.W. Brookfield, C.
van de Bruck and L.M.H. Hall, Phys. Rev. D 74 064028 (2006); F.
Briscese, E. Elizalde, S. Nojiri and S.D. Odintsov, Phys. Lett. B
646 105 (2007)

\bibitem{Nojiri3} S. Nojiri and S.D. Odintsov, Gen. Rel. Grav. 36 1765(2004); Phys.
Lett. B 599  137(2004); G. Cognola, E. Elizalde, S. Nojiri, S.D.
Odintsov and S. Zerbini, JCAP 0502  010(2005); Phys. Rev. D 73
084007(2006); K. Henttunen, T. Multamaki and I. Vilja, Phys. Rev.
D 77 024040(2008); T. Clifton and J. D. Barrow, Phys. Rev. D 72
103005(2005); T. Koivisto, Phys. Rev. D 76 043527 (2007); S.K.
Srivastava, Phys. Lett. B 648 119 (2007); S. Nojiri, S.D. Odintsov
and P. Tretyakov, Phys. Lett. B 651 224 (2007); H. Farajollahi, F. Milani, Mod. Phys. Lett. A 25:2349-2362 (2010)



\bibitem{Setare1} M. R. Setare, M. Jamil, Phys. Lett. B 690  1-4 (2010);
A. C. Davis, C. A.O. Schelpe, D. J. Shaw, Phys.Rev.D80:064016 (92009);
Y. Ito, S. Nojiri, Phys.Rev.D79:103008 (2009);
Takashi Tamaki, Shinji Tsujikawa, Phys.Rev.D78:084028 (2008)

\bibitem{Mota1} D.F. Mota, D.J. Shaw, Phys. Rev. D 75 (2007)
063501; K. Dimopoulos, M. Axenides, JCAP 0506:008
(2005); T. Damour, G. W. Gibbons and C. Gundlach, Phys. Rev. Lett, 64, 123 (1990);
H. Farajollahi, A. Salehi, JCAP 1011:006 (2010);
H. Farajollahi, N. Mohamadi, Int.J.Theor.Phys.49:72-78 (2010);
H. Farajollahi, N. Mohamadi, H. Amiri, Mod. Phys. Lett. A, 25, No. 30 2579-2589 (2010)

\bibitem{Carr} S. M. Carroll, Phys. Rev. Lett. 81 3067(1998); S. M. Carroll, W. H. Press and E. L. Turner, Ann. Rev. Astron. Astrophys, 30, 499 (1992); T. Biswas, R. Brandenberger, A. Mazumdar and T. Multamaki. Phys.Rev. D74 , 063501 , (2006)

\bibitem{Halliwell} J. J. Halliwell, Phys. Lett. B 185, 341 (1987)

\bibitem{Ratra} B. Ratra and P.J.E. Peebles, Phys. Rev. D 37 3406 (1988)

\bibitem{Yokoyama} J. Yokoyama and K.-I. Maeda, Phys. Lett. B 207 31 (1988);
A.A. Coley, J. Ibanez, and R.J. van den Hoogen, J. Math. Phys. 38 5256 (1997); V.D. Ivashchuk, V.N. Melnikov, and A.B. Selivanov, JHEP 0309 059 (2003)

\bibitem{bond} J.R. Bond, G. Efstathiou and M. Tegmark, Mon. Not. Roy. Astron. Soc. 291:L33-L41,(1997)
\bibitem{wang} Y. Wang and P. Mukherjee, ApJ. 650, 1 (2006).
\bibitem{Komatsu} E. Komatsu et al., Astrophys. J. Suppl. 180:306-329 (2009).
\bibitem{Eisenstein} D.J. Eisenstein et al., ApJ. 633, 560 (2005).
\end{thebibliography}
\end{document}